\documentclass[examplefnt,biber,chapter]{nowfnt}

\usepackage[utf8]{inputenc}
\usepackage{tikz}
\usepackage[framemethod=TikZ]{mdframed}
\usepackage{amsfonts,amsmath,amssymb}
\usepackage{biblatex}

\usepackage{graphicx} % Include graphicx package to enable adding images
\usepackage{lipsum} % Include lipsum for dummy text
\usepackage{array}
\usepackage{booktabs}
\usepackage{longtable}
\usepackage{multirow}
\usepackage{enumitem}
\usepackage{tcolorbox}

\DeclareUnicodeCharacter{FFFD}{}

\AtBeginDocument{\mainmatter}

\begin{document}
\chapter{Multi-modal Generative Models in Recommendation System}
\label{ch:mm}

% Preprint ribbon
 \vspace*{-23em}  % Adjust this to move the ribbon closer to the top
{\noindent\small
\linespread{0.8}\selectfont
This is a preprint of Chapter 5 in the upcoming book \textit{Recommendation with Generative Models}
by Yashar Deldjoo, Zhankui He, Julian McAuley, Anton Korikov, Scott Sanner, Arnau Ramisa, Rene Vidal, Mahesh Sathiamoorthy, Atoosa Kasrizadeh, Silvia Milano, and Francesco Ricci.}

% Add space before chapter authors
\vspace{12em}

% Chapter authors
{\noindent\small
\textbf{Chapter Authors\footnote{This work does not relate to the authors positions at Amazon.}:} Arnau Ramisa, Rene Vidal}

\vspace{3em}

\begin{abstract}
The recommendation systems discussed so far typically limit user inputs to text strings or behavior signals such as clicks and purchases,
and system outputs to a list of products sorted by relevance. With the advent of generative AI, users have come to expect richer levels of interactions. In visual search, for example, a user may provide a picture of their desired product along with a natural language modification of the content of the picture (e.g., a dress like the one shown in the picture but in red color).
Moreover, users may want to better understand the recommendations they receive by visualizing how the product fits their use case, e.g., with a representation of how a garment might look on them, or how a furniture item might look in their room. Such advanced levels of interaction require recommendation systems that are able to discover both shared and complementary information about the product across modalities, 
and visualize the product in a realistic and informative way. However, existing systems often treat multiple modalities independently: text search is usually done by comparing the user query to product titles and descriptions, while visual search is typically done by comparing an image provided by the customer to product images. We argue that future recommendation systems will benefit from a multi-modal understanding of the products that leverages the rich information retailers have about both customers and products to come up with the best recommendations.

In this chapter we discuss recommendation systems that use multiple data modalities simultaneously. As we shall see, a key challenge in developing multimodal generative models is to ensure that the features extracted from each modality are adequately \emph{aligned} across modalities, i.e., mapped to nearby points in the embedding space. Since the problem of jointly learning a generative model for each modality and their alignment is extremely difficult~\citep{chen2020generative}, a common approach is to use contrastive learning methods to approximately align the modalities before learning a multimodal generative model. Therefore, in this chapter we will review both contrastive and generative approaches to multimodal recommendation. More specifically, in Section~\ref{sec:intro-mm} we will provide a brief introduction to multimodal recommendation systems, in Section~\ref{sec:discrim-mm} we will review contrastive approaches to multimodal recommendation, and in Section~\ref{sec:gener-mm} we will discuss generative approaches. Finally, in Section~\ref{sec:applications-mm} we will overview various applications of multimodal recommendation systems. Throughout the chapter, we will center the discussion around vision and language models due to the larger volume of work for these two modalities, but we note there is a growing literature of generative recommendation systems that combine other modalities such as audio and text~\citep{vyas2023audiobox}, video and audio~\citep{ruan2023mm}, or even more than two modalities~\citep{wu2023next}.
\end{abstract}

\section{Introduction to Multimodal Recommendation Systems}
\label{sec:intro-mm}

\subsection{Why do we need multimodal recommendation systems?}
\label{sec:mm-why}

Retailers have a lot of information about their customers and the items they sell, including purchase history, customer interactions, product descriptions, product images and videos, and customer reviews. However, existing recommendation systems typically process each data source independently and then combine the recommendation results. For example, text search is typically done by comparing a short user query to product title, descriptions and reviews, while visual search is typically done by comparing an image provided by the customer to product images. Both search approaches produce a list of products sorted by relevance, and current ``multimodal'' systems simply fuse unimodal relevance scores to produce a single list of products from both modalities. In practice, there are many use cases in which such a ``late fusion'' approach may be insufficient for satisfying the needs of the user.

One such use case, known as the \emph{cold start problem}, occurs when new users start using the system, or new products are added to the catalog, hence user behavioral data cannot be leveraged to recommend new products to existing users or existing products to new users.
To alleviate this problem, it is useful to gather diverse information about the items so that preference information can be transferred from existing products or users to new ones. To this end, models that combine information from multiple modalities offer a unique advantage. For example, if a store receives a new product (e.g., a dress), but no purchases have been made yet, we can use the visual similarities between the new dress and existing ones in the store to determine which customers could be interested in it.

Another use case occurs when different modalities are necessary to understand the user request. For example, to answer the request ``best metal and glass black coffee table under \$300 for my living room'', the system would need not only the text query but also an image of the customer's living room in order to find a table that best matches the room. Moreover, answering this customer's question requires reasoning about the appearance and shape of the item in context with the shape and appearance of many other objects, as well as limiting the search by price, which cannot be achieved by searching with either the text or image independently.
Other examples of multimodal requests include an image or audio of the desired item together with modification instructions in text (e.g.,~a dress like the one in the picture but in red, a song like the sound clip provided but in acoustic), or a complementary related product (e.g., a kickstand for the bicycle in the picture, or other movies from the actress talking in the video clip).

A third use case where multimodal understanding becomes crucial is when considering more complex recommendation systems, like those featuring virtual try-on capabilities, or intelligent conversational shopping assistants~\citep{rufus, shopwithai}. To be effective, AI shopping assistants will need to be able to understand the context of previous interactions in the conversation history. 
Let's consider the example of a customer looking for a complete outfit he is planning to wear during the summer in Cairo, to attend the wedding of a friend with traditional tastes. An AI shopping assistant interacting with the customer will have to resort to visual cues to recommend products compatible as an outfit, as well as other customer preferences expressed earlier, the climate in Cairo during the summer, and cultural or dress code norms.

\subsection{Key challenges in designing multimodal recommendation systems}
\label{sec:mm-challenges}

The development of multimodal recommendation systems faces several challenges. 
\begin{itemize}
    \item First, combining different data modalities to improve recommendation results is not simple. Existing systems learn joint representations that capture information that is shared across modalities (e.g., the text query refers to a visual attribute of the product that is visible in the image), but they ignore complementary aspects that could benefit recommendations~\citep{guo2019deep}; e.g., the text mentions inside pockets not visible in the picture, or the image contains texture patterns that are hard to describe precisely in text. Therefore, when learning multimodal representations it is important to ensure adequate alignment of the aspects that need to be aligned, while leaving some flexibility to capture complementary information across modalities as well.
    In general we will want the modalities to compensate for one another and result in a more complete joint representation.
    \item Second, collecting aligned data from multiple modalities to train multimodal recommender systems is significantly more difficult than collecting data for individual data modalities. For example, in the unimodal case one can define positive pairs for contrastive learning via data augmentation, but in the multimodal case such positive pairs often need to be annotated (see Section~\ref{sec:discrim-mm}). In practice, existing annotations may be incomplete for some modalities~\citep{rahate2022multimodal}. For example, visual search with text modification would require examples of an input image, the textual modification, and the modified image, but typically only two of the three are available, e.g., image-caption pairs.
    \item Third, learning a latent space that can be used for generative tasks is often harder than for discriminative tasks, as it typically requires larger datasets and computational resources to be able to adequately learn the data distribution by using more complex losses \citep{chen2020generative}. This challenge is further exacerbated in the case of multimodal data because we need to not only learn a latent representation for each modality but also ensure that these latent representations are adequately aligned.
\end{itemize}

Despite these challenges, multimodal generative models are a promising technology for improving recommendation systems. Indeed, recent literature shows tremendous advances on the necessary components to achieve effective multimodal generative models for recommender systems, including
1) the use of LLMs and diffusion models to generate synthetic data for labeling purposes~\citep{brooks2023instructpix2pix, rosenbaum2022clasp, nguyen2024dataset},
2) high quality unimodal encoders and decoders~\citep{he2022masked, kirillov2023segment}, 
3) better techniques for aligning the latent spaces from multiple modalities into a shared one~\citep{radford2021learning, li2022blip, girdhar2023imagebind},
4) efficient re-parametrizations and training algorithms~\citep{jang2016categorical}, and
5) techniques to inject structure to the learned latent space to make the problem tractable~\citep{croitoru2023diffusion, yang2023diffusion}. 
Once trained, generative recommender systems are more versatile, and can produce better recommendations in more general, open ended tasks.

\subsection{Multimodal recommendation systems covered in this chapter}
\label{sec:mm-summary}

In the remainder of this chapter, we will review both contrastive and generative approaches to multimodal recommendation. In Section~\ref{sec:discrim-mm} we will review contrastive approaches, such as CLIP, which learns to map each modality to a common latent space in which the modalities are approximately aligned. In Section~\ref{sec:gener-mm} we will discuss generative approaches, such as ContrastVAE, which learns a probabilistic embedding from each modality to a common latent space where modalities are approximately aligned, and DALL-E 2, Stable Diffusion, LLAVA and multimodal LLMs, which learn to generate image recommendations given an input text prompt.

\section{Contrastive Multimodal Recommendation Systems}
\label{sec:discrim-mm}

As discussed in Chapter~4.3.1, many recommendation approaches like~\cite{du2022amazon} rely on learning an embedding of the data such that similar items are close to each other in the embedded space. In the case of multimodal data, a natural approach to learning a \emph{multimodal embedding} would be to learn one embedding per modality, as done by~\citep{he2020momentum, 10.5555/3495724.3497510, chen2020simple, Caron_2021_ICCV} for images, or~\citep{saeed2021contrastive, won2020data, wang2022towards} for audio, and then concatenate such \emph{unimodal embeddings}. Such an approach is adequate when different modalities capture complementary aspects of an item. However, as discussed in Section~\ref{sec:mm-challenges}, when different modalities capture related aspects of an item, unimodal embeddings need to be adequately aligned to ensure that similar items are close to each other in the multimodal embedded space. For example, to ensure that the embedding of a textual description of a product is close to the embedding of an image of the same product we need to learn both embeddings with that constraint in mind, which often requires large amounts of aligned training data (e.g., text-image pairs).
One way to address this challenge is to first learn an alignment between data modalities and then learn a generative model on \emph{aligned} representations. Hence, in this section we will focus on the problem of learning aligned representations across multiple modalities.

A popular approach to learning aligned representations is contrastive learning~\citep{gutmann2010noise} which, for a pair of data points from different modalities, minimizes a loss that encourages their embeddings to be close when the points are similar (positive pairs), and far when the points are very different (negative pairs). In the single modality setting, positive pairs are generated by simply altering one sample (e.g., slightly shifting the image, flipping the image, transforming it to grayscale). In the multimodal setting, however, it is hard to generate a corresponding positive pair in the other modality via simple augmentation strategies. Instead, positive pairs are typically obtained by labeling similar pairs in a coarse-grained or fine-grained manner. Coarse-grained labels (e.g., a pair of an image and a caption) are easier to obtain, but they may not be sufficiently discriminative. Fine-grained labels (e.g., a bounding box for each object in the image and the corresponding word in the caption) are harder to obtain, but they provide more detailed correspondences between image regions and words in the caption.

\subsection{Contrastive Language-Image Pre-training (CLIP)}
Contrastive Language-Image Pre-training (CLIP)~\citep{radford2021learning} is one of the most popular contrastive learning approaches to multi-modal pre-training. The main idea behind CLIP is that coarse labels in natural language have a sufficient degree of supervision to enable the learning of general concepts, while being much easier to scale using internet data. Indeed, the authors of CLIP found that trying to predict the exact words, as previous works had done, led to very hard training objectives that converged very slowly, due to the variety of ways in which the same information can be conveyed.
Therefore, they proposed to use coarse labels, i.e., to pair an entire image with a caption.

\begin{figure}[tb]
\centerline{\includegraphics[width=0.7\columnwidth,keepaspectratio]{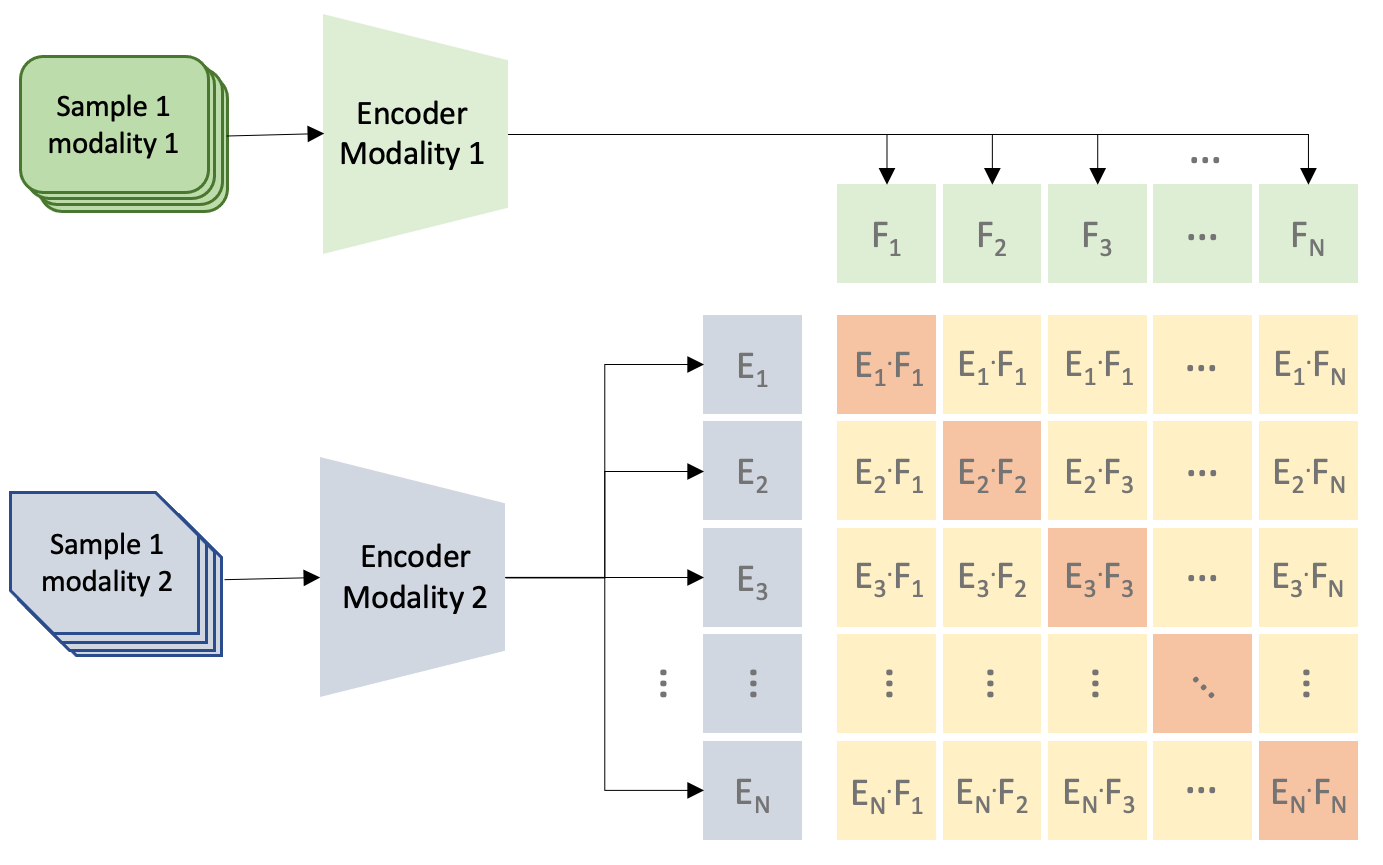}}
\caption{Contrastive pre-training used to train models such as CLIP. For each minibatch, the positive (diagonal) and negative (off-diagonal) pairs are used to compute the loss.}
\label{fig:clip-architecture}
\end{figure}

Figure~\ref{fig:clip-architecture} shows CLIP's model architecture, which consists of two towers, an image encoder and a text encoder, that project an input image-text pair to a shared embedding space. Semantically equivalent image-text pairs should be projected to the same point in the embedding space, and unrelated image-text pairs should be projected to far apart points. This is achieved by computing the cosine similarity for all possible image-text pairs in a training minibatch, and applying a symmetric cross-entropy loss over the rows and columns of the similarity matrix.

To be effective, CLIP was trained on a large dataset of 400 million image and text pairs, obtained from downloading images and their associated \textit{alt-text} (text to be displayed in place of an image that fails to load) from the Internet. The dataset was curated to guarantee coverage of concepts by balancing the word occurrences, and filtering with methods like making sure each text included at least one word from a pre-defined list obtained from Wikipedia data to remove noisy or irrelevant image-text pairs. 
While aligning crawled Internet images and their alt-text is bound to find many irrelevant or misleading examples, dataset curation techniques to improve the quality of training samples~\citep{cao2023less, fan2023improving}, or to reduce harmful or undesired examples~\citep{bansal2023cleanclip, yu2023devil} have proven useful to improve results. Furthermore, scaling up the datasets to billion-scale~\citep{schuhmann2022laion} has shown that noisy examples not previously removed can be cancelled by overwhelming numbers of positive ones, resulting in better overall performance~\citet{jia2021scaling}.

The simple idea behind CLIP demonstrated to scale very well and achieved state-of-the-art in many zero-shot benchmarks. For example, it obtained impressive zero-shot classification and retrieval results~\citep{novack2023chils, baldrati2023zero, hendriksen2022extending}, and has been successfully fine-tuned to a multitude of tasks, such as object detection~\citep{Gu2021OpenvocabularyOD}, semantic segmentation~\citep{zhou2023zegclip} or action recognition~\citep{huang2024froster}. The same contrastive alignment objective has also been used between other modalities, including audio and images ~\citep{cheng2020look}, tables and images~\citep{hager2023best}, tables and medical images~\citep{huang2023multimodal}, and with multiple modalities at the same time~\citep{girdhar2023imagebind}. 
The datasets used for pre-training these models are typically composed of data scrapped from the Internet (e.g.,~pairs of images and alt-text), generated as a byproduct of another process (e.g.,~e-commerce purchases~\citep{Chen2023BeyondSL}, robot sensor logs~\citep{huang2021cross}), or automatically generated by an existing ML model (e.g. speech in audio or video~\citep{zhang2021contrastive}, human poses in images computed with OpenPose~\citep{cao19openpose}). 

The generalization ability of CLIP and similar Vision-Language Models (VLM) greatly benefited from scaling the training in model size, batch size, and dataset size~\citep{pham2023combined, cherti2023reproducible}. 
Researchers have also studied how preferring adaptation of the text branch over the language branch affected results~\citep{zhai2022lit}. Furthermore, many approaches have been proposed to improve the semantic accuracy of the resulting models~\citep{li2023multimodal}, such as loss functions to improve the image and text encoders~\citep{he2022masked, shen2022k}, or to encourage desirable properties such as multilanguage understanding, interpretability and fairness in the embedding space~\citep{chen2023stair, carlsson2022cross, dehdashtianfairerclip}.
Other interesting improvements include training better encoders with additional loses like image masking~\citep{he2022masked} and Triple Contrastive Learning~\citep{yang2022vision}, or enhancing the text with Wikipedia definitions of entities~\citep{shen2022k}.

\subsection{Other Contrastive Pre-training Approaches}
Other approaches have looked into novel architecture designs and novel losses to further improve results. Align BEfore Fuse (ALBEF)~\citep{li2021align}, for example, uses a multimodal encoder to combine the text and image embeddings generated from the unimodal encoders, and propose two additional objectives to pre-train a model in addition to the Image-text contrastive (ITC) learning: masked language modeling (MLM) to predict masked words on the unimodal text encoder, and image-text matching (ITM) to classify if a pair of image and text match or not. 
The authors also introduce \textit{momentum distillation}, where a moving average version of the model weights provides pseudo-labels in order to compensate for the potentially incomplete, or wrong, text descriptions in the noisy web training data. Using their proposed architecture and training objectives, ALBEF obtains better results than CLIP in several zero-shot and fine-tuned multimodal benchmarks, despite using orders of magnitude less images for pre-training. In a subsequent work, \citet{li2022blip} replace the multimodal encoder by cross-attention layers to the text tower to model vision-language interactions, and replace the MLM loss by a Language Modelling (LM) loss that trains the model to maximize the likelihood of a generated caption given an image.

Finally, other works explore how to bring more modalities into alignment. \citet{girdhar2023imagebind} propose ImageBind, an approach to learn an aligned embedding across six different modalities, including text, audio, image, depth, thermal and Inertial Measurement Unit (IMU) data. Instead of requiring paired data for all modalities, they only rely on readily available paired data between image and other modalities (e.g., web scale text-image data, audio for a video clip or depth in RGBD images). All modality encoders use transformer networks and the joint model is learned using the InfoNCE loss. 

These contrastive multimodal models can then be used in multimodal recommendation systems such as~\citep{sevegnani2022contrastive, alpay2023multimodal, wu2023next}. They are also used to initialize the weights of generative multimodal systems, that will make the generative training much more tractable.

\section{Generative Multimodal Recommendation Systems}
\label{sec:gener-mm}

Despite their advantages, purely contrastive recommendation systems often suffer from data sparsity and data uncertainty~\citep{wang2022contrastvae,lin2023contrastive}. For example, users may provide reviews for very few items and some of them may have errors. Generative models address these issues by imposing structure on the data generation process, e.g., by using latent variable models, and by adequately modeling uncertainty. Moreover, generative models allow for more complex recommendations, e.g., those involving image generation. 

In this section, we will survey generative recommendation systems that utilize multiple modalities in order to better understand the user, or provide the recommendations. Depending on how the generative models are designed and learned, we will distinguish between three types of models: Generative Adversarial Networks (see Section \ref{sec:gan-mm-rec}), Variational AutoEncoders (see Section \ref{sec:vae-mm-rec}), and Diffusion Models (see Section \ref{sec:diff-mm-rec}). All these three types of models posit the existence of a latent variable $Z$ (continuous or discrete) such that the distribution of the data $X$ (e.g., image and text) can be written as
\begin{equation}
\label{eq:latent-variable-model}
    p(X) = \int p(X\mid Z) p(Z) dZ. 
\end{equation}
The main differences among these models are how the prior $p(Z)$ and posterior $p(X\mid Z)$ are defined and parametrized with deep networks, and what losses are used to learn the network weights from data. The following subsections describe each one of these models in more detail, how network architectures are modified to accomodate multimodal data, and how these models are used for building recommendation systems.

\subsection{Generative Adversarial Networks for Multimodal Recommendation}
\label{sec:gan-mm-rec}

Proposed by~\citet{goodfellow2014generative}, Generative Adversarial Networks (GANs) are an innovative approach to learning a distribution from multimodal data. GANs have been used in various recommender systems, including collaborative filtering \citep{wei2023multi} and content-based retrieval \citep{tautkute2021want}. In this subsection, we  will briefly summarize the basic formulation of GANs for unimodal data, show how it can be extended to multimodal data, and discuss adaptations of GANs for collaborative filtering and content-based retrieval.

\begin{figure}[tb]
\centerline{\includegraphics[width=0.8\columnwidth,keepaspectratio]{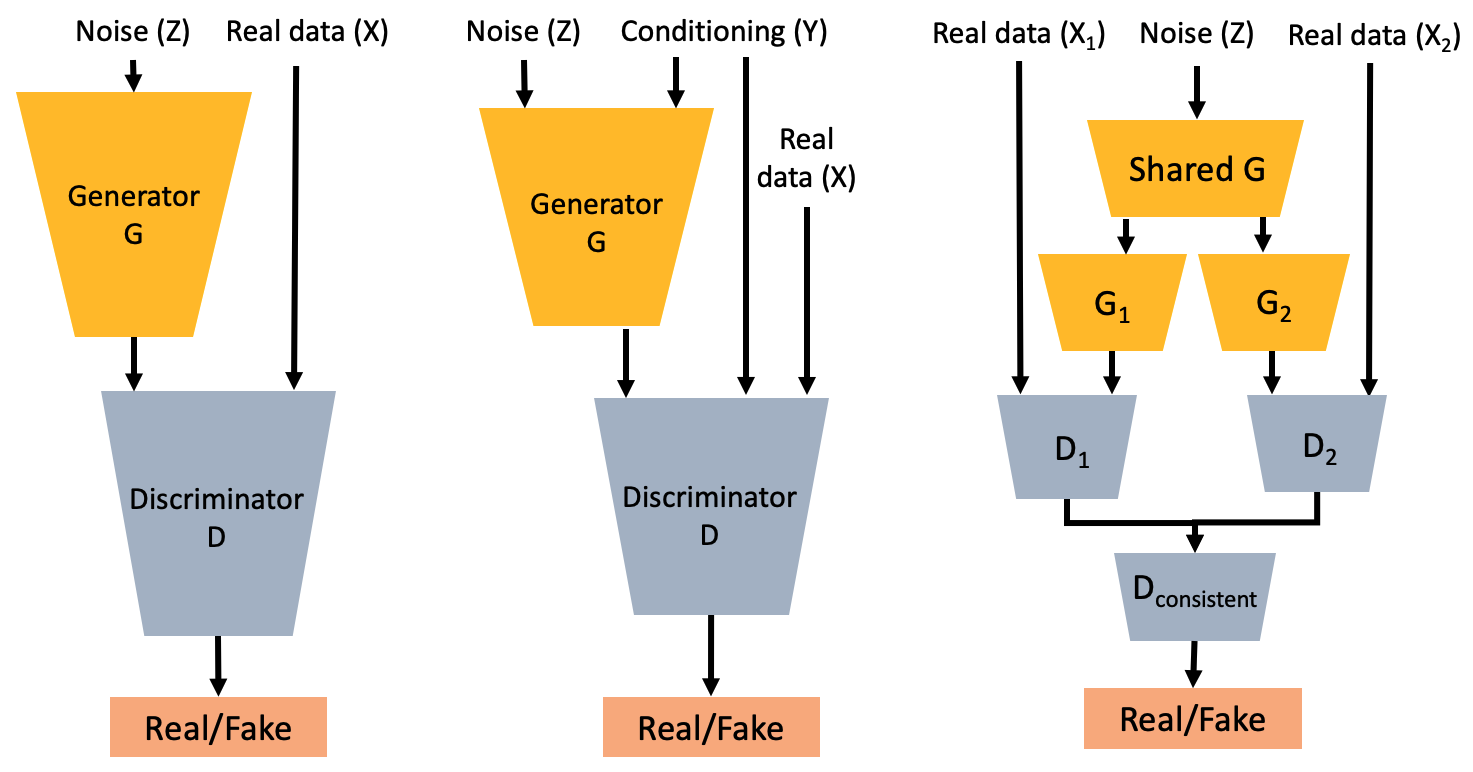}}
\caption{Generative Adversarial Networks (GANs) are composed of a generator $G$ that generates a data point (e.g. an image) from a latent variable $Z$, and a discriminator $D$ which tries to determine if a data point is fake (synthesized by the generator) or real. Left: Standard GAN architecture for unconditional generation. Center:  Conditional GAN architecture. Right: Multimodal GAN architecture.}
\label{fig:GAN}
\end{figure}

\paragraph{Unimodal GANs} 
As discussed in Section 3.5.5, GANs are a class of latent variable models in which a generator maps a latent variable $Z$ to a sample data point $X$, while a discriminator $D$ decides whether its input is real or generated (see Fig. \ref{fig:GAN} left). More specifically, GANs learn a probability distribution as in Eq.  \eqref{eq:latent-variable-model}, where $p(Z)$ denotes the prior which is typically assumed to be a standard Gaussian (continuous) or categorical (discrete) and $p(X\mid Z)$ denotes the posterior. What is unique in GANs is that the posterior $p(X\mid Z)$ is not modeled explicitly (e.g., as a Gaussian). Instead, the posterior is represented by a generator $G: \mathcal{Z}\to\mathcal{X}$ that produces samples $G(Z)$ from $p(X\mid Z)$ without having to represent $p(X\mid Z)$. Then, a second component of a GAN is the discriminator $D: \mathcal{X}\to \{0,1\}$, which is designed to discriminate real from generated images, i.e., $D(X)=1$ if $X$ is real and $D(X)=0$ if $X$ is generated. The generator and discriminator are then jointly learned from samples of $p(X)$ by optimizing a min-max objective 
\begin{equation}
\label{eq:GAN-objective}
\min_{G} \max_{D} 
\mathbb{E}_{X \sim p(X)}
[\log D(X)]
+
\mathbb{E}_{Z \sim p(Z)}
[\log (1-D(G(Z))]
\end{equation}
in which the generator $G$ tries to generate samples that fool the discriminator $D$, while $D$ tries to discriminate between real samples $X\sim P(X)$ and generated samples $G(Z)$, $Z\sim P(Z)$. 

One advantage of GANs is that sampling is straightforward: all we need to do is to sample $Z$ (e.g., categorical or standard Gaussian) and pass it through the generator to produce $X$. Another advantage is that, in the ideal case in which $G$ and $D$ have infinite capacity, one can show that the optimal discriminator $D^*$ can't tell true from generated (i.e., $D^*(X) = 1/2$) and the optimal generator $G^*$ is such that the distribution of the generated data $G^*(Z)$ matches the distribution of the true data $X$.
In practice, $D$ and $G$ are parametrized with neural networks, and the expectation in the min-max objective is computed as the average over samples. As a consequence, while there is no guarantee that GANs learn the true distribution of the data, in the case of images it has been empirically shown that GANs produce high quality generations.

Despite these advantages, GANs also suffer from some limitations. One of them is the issue of mode collapse, which happens when the generator produces samples that are not representative of the full data distribution, such as generating only the most likely outputs, or a specific output that fools the discriminator~\citep{zhang19vanishing}. GANs also suffer from training instabilities due to the nature of the min-max objective optimized by the generator and discriminator networks. For example, \citep{arjovsky17waserstein} show that a small change in one network leads to major adjustments in the other, which can result in destabilizing the learning process and failing to converge. Moreover, gradient vanishing problems happen when one network dominates the other, e.g., when the discriminator becomes very accurate and produces a loss with little gradient information for the generator~\citep{Su2018GANQPAN, Chakraborty_2024}.

\paragraph{Multimodal GANs}
The vanilla GAN formulation discussed so far assumes that $X$ is generic, i.e., $X$ can be unimodal or multimodal. In principle, we could use such a vanilla formulation to learn generative models for multimodal data. However, doing so may require collecting, annotating and aligning very large datasets and using them to train a very complex multimodal generator. In practice, it may be preferable to design specialized models that leverage existing unimodal generators, such as models that can generate one modality conditioned on another, or models that can ensure adequate alignment across modalities.

Conditional GANs (see Fig. \ref{fig:GAN} center) generate data for data modality $X$ conditioned on another modality $Y$, such as the product type, a textual description of a product, an image mask, etc. In this case, the goal is to learn a conditional model of the form
\begin{equation}
\label{eq:conditional-latent-variable-model}
    p(X \mid Y ) = \int p(X\mid Z, Y) p(Z) dZ. 
\end{equation}
To model $p(X \mid Z, Y)$, the generator must take both $Z$ and $Y$ as inputs to generate samples $G(Z,Y)$. Likewise, the discriminator $D(X,Y)$ must also depend on the conditioning variable $Y$. Different modalities can be used for the condition; examples are class-conditioning~\citep{MirzaO14conditional}, conditioning on an input image~\citep{isola17pix2pix}, or using a latent code vector~\citep{chen16infogan}. \citet{huang2022multimodal} proposed an approach to allow conditioning on multiple input modalities (e.g., text, sketch, segmentation mask) to generate new images. This allowed very fine-grained control of the generated image layout and content. \citet{ziegler2022multi} also use conditioning on multi-modal clinical tabular data for the generation of realistic 3D medical images.

Alternatively, we may want to generate multiple data modalities. For the sake of simplicity, assume that the data is composed of two modalities $X=(X^1,X^2)$. We can design a multimodal generator that leverages unimodal generators by assuming that $X^1$ and $X^2$ are conditionally independent given $Z$. Under these assumptions, the model in Eq. \eqref{eq:latent-variable-model} factorizes as the product of two unimodal models because
\begin{equation}
\label{eq:MM-latent-variable-model}
    p(X^1,X^2) = \int p(X^1,X^2\mid Z) p(Z) dZ = \int p(X^1\mid Z) p(X^2\mid Z) p(Z) dZ. 
\end{equation}
Therefore, we can use one generator per modality, $X^1 = G^1(Z)$ and $X^2 = G^2(Z)$, to represent $p(X^1\mid Z)$ and $p(X^2\mid Z)$, respectively. Note, however, that the latent representation $Z$ must be shared to ensure alignment across modalities. Alternatively, we may want the generators $G_1$ and $G_2$ to have a shared backbone that then splits into separate branches for each modality (see Fig. \ref{fig:GAN} top right). For example, \citet{zhu2024consistent} use a StyleGAN backbone with three modality specific branches. 

Regarding the design of multimodal discriminators, we note that the discriminator should take generated data for both modalities and compare it with the true data for both modalities. To leverage pre-trained discriminators for each specific modality, say $D_1$ and $D_2$, we could simply fuse the predictions of unimodal discriminators. Alternatively, we could fuse intermediate features from unimodal discriminators and have a simple discriminator $D_{consistent}$ predict whether the data is real or fake from the fused representation of both modalities (see Fig. \ref{fig:GAN} bottom right).
\citet{zhu2024consistent} also use two types of discriminators: \emph{fidelity discriminators} are unimodal discriminators that assess the quality of an individual data modality, while \emph{consistency discriminators} judge whether two modalities are consistent with each other.

\paragraph{Multimodal GANs for collaborative filtering} As discussed before, a natural approach to building multimodal recommendation systems is to incorporate multiple modalities when learning a latent representation of items and/or users. However, existing multimodal representation learning methods lack robustness to scarce labels for user-item interactions. Self-supervised learning methods address this problem by exploiting supervisory signals in unlabeled data, e.g., by using data augmentation. However, a key challenge is generating augmentations that are consistent across multiple modalities. Recent work \citep{wei2023multi} proposes an adversarial multi-modal self-supervised learning paradigm in which a generator proposes collaborative relations which are then vetted by a discriminator. In addition, \citet{wei2023multi} propose a cross-modal contrastive learning framework for preserving inter-modal semantic commonality and user preference diversity. On the other hand, GANs have been used to model and improve user-item interaction data. For example, \citet{gao2021recommender} review several works that use GANs to mitigate noise and perform informative sample selection in user preferences data, and to synthesize new samples through data augmentation.

\paragraph{Multimodal GANs for fashion recommendation}

Due to its visual nature, GANs have found an application area in fashion-related tasks such as compatible outfit generation, virtual try-on, or product search. 

 \citet{liu2021clothing} tackle clothing compatibility learning. Given an image of a clothing product, and a target product category name, the proposed method generates an image of a compatible item with a GAN. 
 A compatibility matrix representing a style space is used to condition the GAN and
 make sure the generated item is compatible with the input one.
 The style space is learnt using triplets of anchor with compatible and incompatible clothing items, and additional losses for feature matching, reconstruction, and a discriminator loss.
\citet{zhou2023bc} propose a method to generate multiple options for compatible clothing simultaneously, with attention to diversity. 
They use a style embedding discriminator to provide supervision to the generator through a binary real/fake classification loss, and a compatibility discriminator that uses a contrastive loss. They also include a diversity loss to ensure variety in the generated items.

For virtual try-on, an input garment image, and a target image with a person onto which the garment has to be placed are used. Generators often use several modalities derived from the target image, such as the person mask and a pose image, to condition the generation.
For example, \citet{liu2019swapgan} take conditional and reference images and transfer the clothing from the person in the conditional image to the one in the reference image. For that, they use a pose map, segmentation map, mask map and head map, derived from the input images. They combine three generators and a discriminator to have a single system for clothing transfer. 
Similarly, \citet{pandey2020poly} use segmentation masks, pose estimation and clothing parsing (i.e.,~detecting all clothing in a picture) to transfer a reference garment in a white background image to a person in a model image. Their proposed system combines multiple tasks previously done by different networks into a single architecture.

\citet{tautkute2021want} use GANs for query expansion (i.e.,~augmenting or reformulating the user query to improve retrieval results) on a multimodal fashion product retrieval scenario. Instead of combining or fusing the text and image representations, as is commonly done in multimodal search, they generate an image of the desired product, and then use it for visual search. The authors trained their image generation network using both a discriminator and a triplet loss to make sure the generated image is not too close to the original query image.

\subsection{Variational AutoEncoders for Multimodal Recommendation}
\label{sec:vae-mm-rec}

Proposed by \citet{Kingma2014}, Variational AutoEncoders (VAEs) have been used as a key component of various recommendation systems, including collaborative filtering with implicit feedback \citep{liang2018variational}, collaborative filtering with side information \citep{karamanolakis2018item}, and content-based retrieval \citep{yi2021cross}. In this subsection, we  briefly summarize the formulation of VAEs for unimodal data, and show how it can be extended to multimodal data, including adaptations of VAEs for collaborative filtering and content-based retrieval.

\begin{figure}[tb]
\centerline{\includegraphics[width=0.9\columnwidth,keepaspectratio]{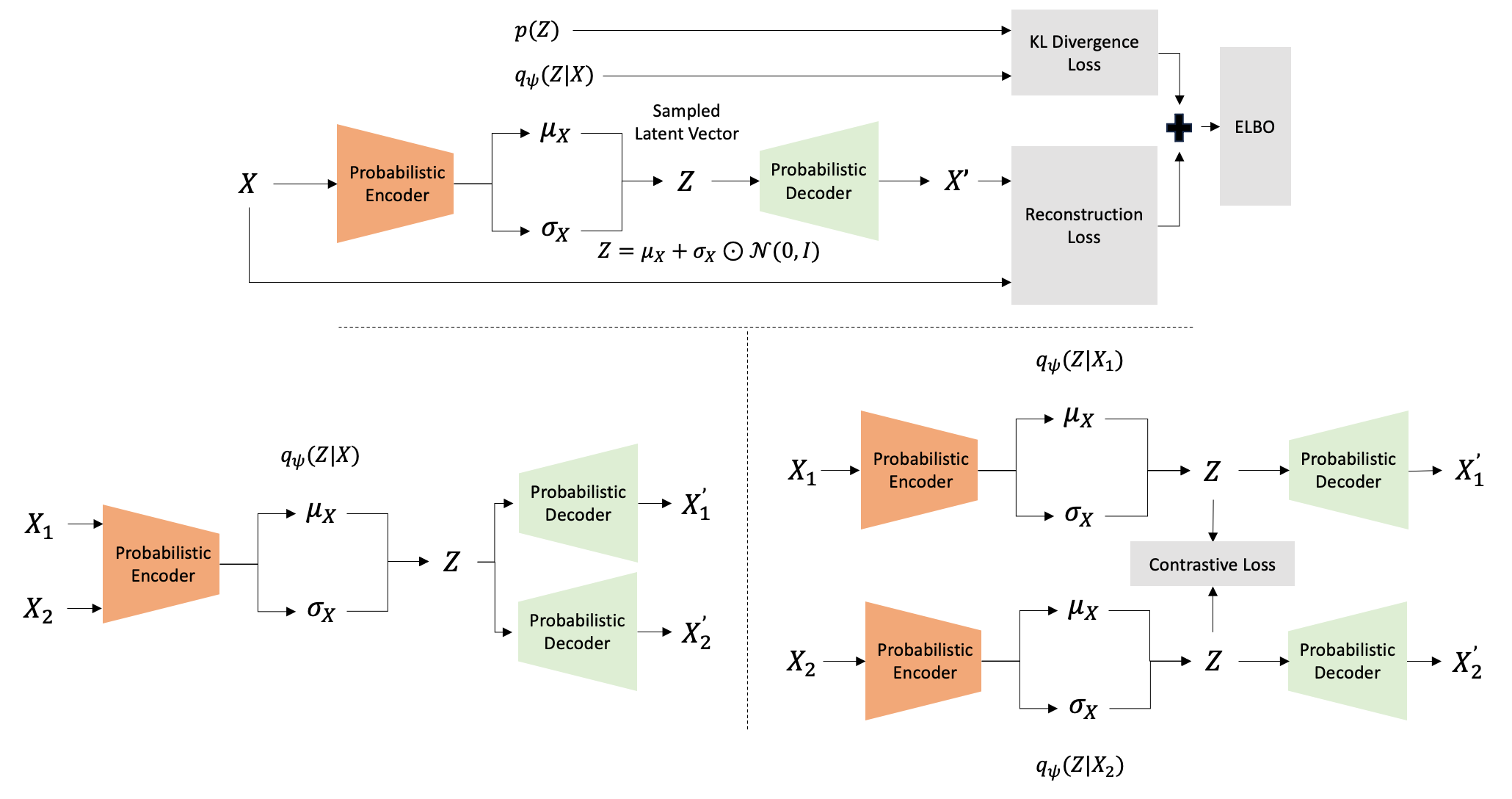}}
\caption{A Variational AutoEncoder (VAE) is composed of a probabilistic encoder that maps an input data point to a latent distribution from which a latent vector is sampled, and decoded with a probabilistic decoder with the objective of reconstructing the original input. Top: Standard VAE architecture for unconditional generation. Bottom left: Multimodal VAE architecture with a shared encoder and two unimodal decoders. Bottom right: Contrastive VAE architecture with two unimodal encoders and decoders and a contrastive loss for aligning the latent spaces.}
\label{fig:VAE}
\end{figure}

\paragraph{Unimodal VAEs}
As discussed in Section~3.5.1, VAEs are a class of latent variable models in which a \emph{probabilistic encoder} maps the input data $X$ to a latent variable $Z$, and a \emph{probabilistic decoder} maps $Z$ back to $X$ (see Fig. \ref{fig:VAE} top). More specifically, VAEs learn a probability distribution $p(X)$ for data $X$ (e.g., image or text) by positing the existence of a latent variable $Z$ (continuous or discrete) such that $p(X) = \int p_\theta(X\mid Z) p(Z) dZ$. The prior $p(Z)$ is typically assumed to be a standard Gaussian (continuous) or uniform (discrete). The posterior $p_\theta(X\mid Z)$ is typically assumed to be Gaussian or categorical and is implemented by a neural network (encoder) that maps $Z$ to the parameters of $p_\theta(X\mid Z)$, e.g., its mean $\mu_\theta(Z)$. The posterior $p_\theta(Z\mid X)$ is typically intractable and thus approximated by a simpler distribution $q_\psi(Z\mid X)$ (e.g., Gaussian or categorical) implemented by another neural network (encoder), which maps $X$ to, e.g., the mean $\mu_\psi(X)$. The weights of the encoder-decoder pair are learned by maximizing a lower bound for the log likelihood $\log(p(X))$, known as the Evidence Lower Bound (ELBO), 
\begin{equation}
\label{eq:ELBO}
    \mathcal{L} = \mathbb{E}_{Z\sim q_\psi(Z\mid X))} \Big [ \log p_\theta (X \mid Z) - KL (q_\psi(Z\mid X) \mid\mid p_\theta(Z\mid X)) \Big ]
\end{equation}
which is the sum of a reconstruction term $\log p_\theta (X \mid Z)$ and a regularization term $KL(q_\psi(Z\mid X) \mid\mid p(Z))$. The variable $Z$ is then used for downstream recommendation tasks.

\paragraph{Multimodal VAEs}
In the case of multimodal data, say $X=(X^1,X^2)$ consists of both image and text, we can still use the VAE model described so far. However, as we argued in the case of GANs, doing so may require designing a very complex decoder. A better approach is to design multimodal VAEs that leverage unimodal VAEs. For example, as we did in \eqref{eq:MM-latent-variable-model}, we can assume that $X^1$ and $X^2$ are conditionally independent given $Z$, i.e., $ p_\theta(X^1,X^2\mid Z)  = p_{\theta_1}(X^1\mid Z) p_{\theta_2}(X^2\mid Z)$, so that we can use one decoder per modality. However, since the latent space $Z$ is shared, this requires the design of a shared encoder $q_\psi(Z\mid X^1, X^2)$. Fig. \ref{fig:VAE} (bottom left) shows the design of such a multimodal VAE with a single probabilistic encoder and two modality specific decoders.

To leverage modality specific encoders and decoders pretrained on large datasets, two families of approaches have been proposed. The first family approximates $q_\psi(Z\mid X^1, X^2)$ with a product of experts \citep{wu2018multimodal}, a mixture of experts \citep{shi2019variational}
 or a mixture of products of experts \citep{sutter2020multimodal}, allowing one to fuse multiple unimodal encoders into a multimodal one. The second family, partitions
the latent space per modality, i.e., $Z=(Z^1,Z^2)$, and assume that $q_\psi(Z \mid X) = q_\psi(Z^1\mid X^1) q_\psi(Z^2 \mid X^2)$ and $p_\theta(X \mid Z) = p_\theta(X^1\mid Z^1) p_\theta(X^2 \mid Z^2)$. However, doing so reduces the entire model to two independent VAEs, one per modality, which defeats the purpose of having a multimodal model. ContrastVAE \citep{wang2022contrastvae} addresses this issue by adding a contrastive loss to the ELBO objective, the InfoNCE loss \citep{oord2018representation}, which aligns the latent spaces of the two modalities. Experiments in ~\citet{wang2022contrastvae} show that ContrastVAE improves upon purely contrastive models by adequately modeling data uncertainty and data sparsity, and being robust to perturbations in the latent space.

\paragraph{Multimodal VAEs for collaborative filtering} 
Traditional VAEs for recommendation systems are unimodal in nature as they aim to model user ratings. For example, \citet{liang2018variational} extends VAEs to collaborative filtering for implicit feedback by using a multinomial likelihood conditional likelihood. 
However, such models often use the standard Gaussian as a prior, which has been shown to give poor latent representations [16]. \citet{karamanolakis2018item} extend VAEs to collaborative filtering with side information. Their key contribution is to replace the standard Gaussian prior in the latent space of the VAE (which is user-agnostic) by a prior that incorporates multimodal user preferences (e.g., user reviews and ratings). The resulting VAE achieves around 30\% relative improvement in ranking metric with respect to standard VAEs for collaborative filtering.

\paragraph{Multimodal VAEs for content-based retrieval} 
\citet{yi2021cross} proposes a multimodal VAE for content-based retrieval. The proposed approach takes three modalities (music, video, and text), maps each modality to a separate latent space using modality-specific encoders, and then aligns these latent spaces via cross-modal generation. More specifically, the video and text modalities are first fused via a product-of-experts model and the fused representation is passed through a cross-modal decoder that generates music. Conversely, the encoding of music is passed through another crossmodal decoder that generates the visual representation. The resulting representation is trained in 150000 video clips of 3000 different music backgrounds and used to build a music recommendation system. 

\paragraph{Graph VAE for Multimodal Recommendation}
In many applications, multimodal data are better represented by a graph. For example, the graph nodes can be items with hand-crafted or learned features from all modalities, while the graph edges can represent item-item similarities. If we want to learn a VAE for the graph and its features, the encoder needs to be able to process a graph as an input and the decoder needs to generate a graph as an output. Graph neural networks (GNNs) are specialized architectures for processing graphs and can be used as both encoders and decoders. The latent space can be a Gaussian vector, as before, or a graph with one Gaussian vector per node. The resulting model is known as a Graph VAE 
 or GVAE for short \citep{kipf2016variational}, and has been used in various recommendation systems.

One example is the work if
\citet{zhou2024disentangled}, which proposes a Disentangled Graph Variational AutoEncoder (DGVAE) for interpretable multimodal recommendation. DGVAE harnesses contrastive pretraining approaches to map multimodal data to a common space in which user-to-item and user-to-word similarities are used to build an item-to-item graph, which is processed by a GNN. Mutual information maximization is used to regularize the learning objective. Experiments show significant improvements in retrieval performance, especially in terms of the interpretability of the recommendations.

Another example is the work of 
\citet{chattopadhyay2023learning}, which uses a conditional GVAE to generate decoration recommendations for a room given its type (e.g., bedroom) and its layout (e.g., room elements such as floor and walls). A graph is used to represent both room and furniture layouts, e.g., the nodes capture attributes such as the location, orientation and shape of room and furniture elements, while the edges capture geometric relationships such as relative orientation. Their GVAE then generates a furniture graph, e.g., a collection of furniture items such as bed and night stand that is consistent with the room type and layout, which is then rendered to obtain images of the decorated room. Experiments on the 3D-FRONT dataset show that their method produces scenes that are diverse and adapted to the room layout.

\subsection{Diffusion Models for Multimodal Recommendation}
\label{sec:diff-mm-rec}

Diffusion models~\citep{sohl2015deep} have recently emerged as the state-of-the-art approaches for generation of images and other data modalities. They take inspiration from stochastic differential equations \citep{Feller1949}, dynamical systems \citep{anderson1982reverse} and non-equilibrium thermodynamics~\citep{sohl2015deep} to learn highly complex data distributions.
Several works have used multimodal diffusion models for recommendation. For example, \citet{zihao23diffurec} use diffusion models for sequential recommendation, and \citet{seyfioglu2024diffuse} propose a fast diffusion model for virtual try-on that takes a picture with a human model and a white background catalog picture of a garment as input, and places the garment on the model. In this subsection we will discuss the basic diffusion model architecture, its extensions to multimodal data, and applications in content generation.

\begin{figure}[tb]
\centerline{\includegraphics[width=0.9\columnwidth,keepaspectratio]{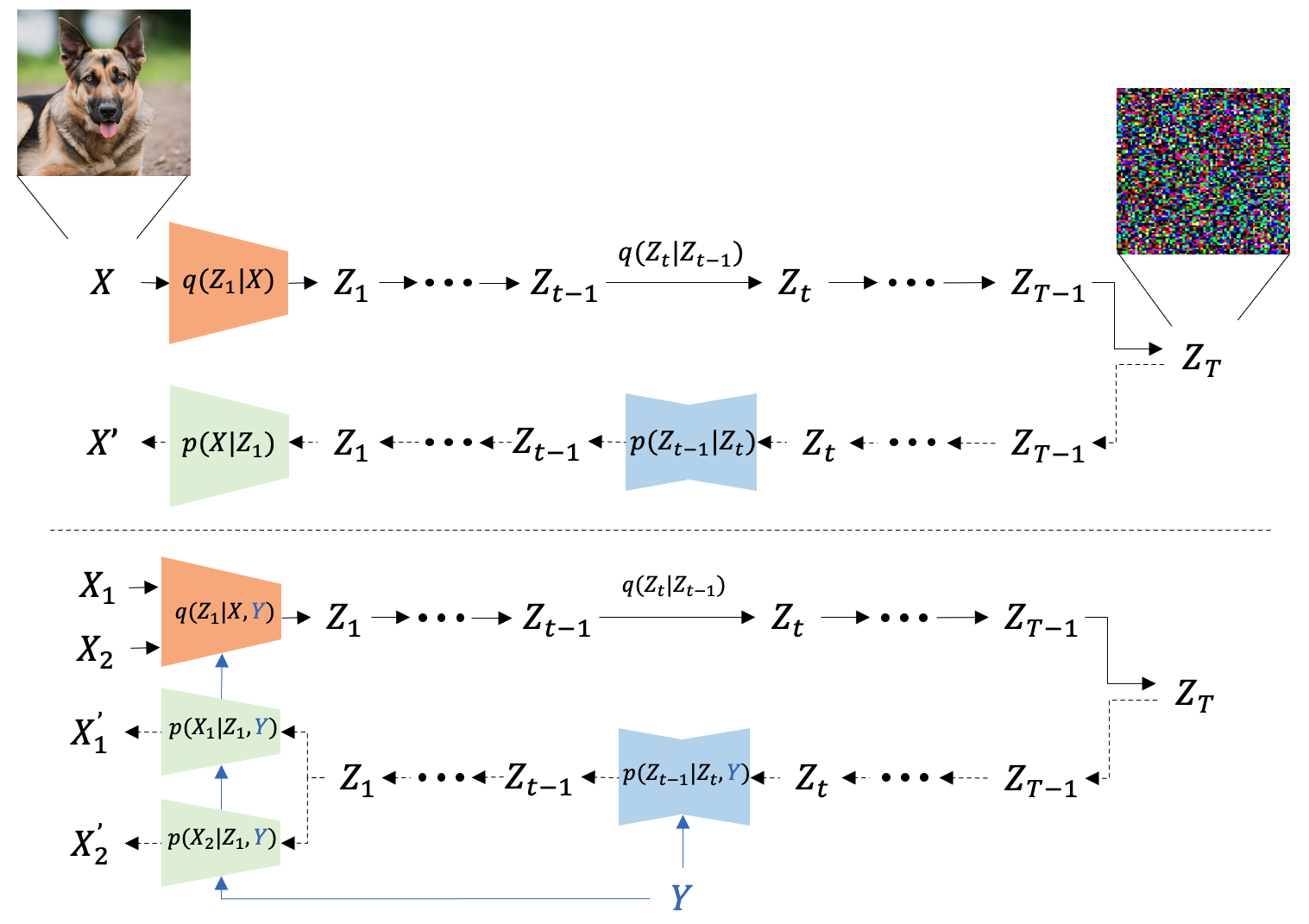}}
\caption{A diffusion models consists of a forward process, which iteratively corrupts an input data sample until it becomes Gaussian noise, and a reverse process, which reconstruct the original data from white noise. Top: Latent diffusion model architecture for unconditional generation. The standard diffusion model architecture is obtained by removing the encoder $p$ and the decoder $q$. Bottom: Conditional multimodal diffusion model with a shared encoder and two unimodal decoders. An unconditional multimodal model is obtained by simply removing the condition on $Y$.}
\label{fig:diffusion-architecture}
\end{figure}

\paragraph{Unimodal diffusion models}
The main idea behind a diffusion model is to generate a new image by sampling a random Gaussian vector and transforming it via multiple denoising steps. This is done by defining a forward diffusion process\footnote{The forward process is also called corruption or noising process, while the reverse process is also called restoration or denoising process.} $X \to Z_1 \to Z_2 \to \cdots \to Z_T$ that iteratively maps a data sample $X$ to Gaussian noise $Z_T$, and a reverse diffusion process $Z_T \to Z_{T-1} \to \cdots \to Z_1 \to X$ that recovers the original data from noise. More specifically, diffusion models are VAEs with a sequential latent space $Z=(Z_1,\dots, Z_T)$. The VAE encoder $q(Z \mid X) = q(Z_1\mid X) \prod_{t=2}^T q(Z_t \mid Z_{t-1})$ assumes that $Z\mid X$ is Markovian with Gaussian transition probabilities $q(Z_t \mid Z_{t-1})$. The VAE decoder $p(X \mid Z) = p(Z_T) p(X \mid Z_1) \prod_{t=2}^T p(Z_{t-1} \mid Z_{t})$ assumes that $Z$ is Markovian with Gaussian transition probabilities $p(Z_{t-1}\mid Z_t)$ and $p(Z_T)$ a standard Gaussian. The transition probabilities are parametrized with deep networks whose parameters are learned by maximizing the ELBO objective in \eqref{eq:ELBO}. Once trained, the model can generate high-quality original data examples by sampling from the noise distribution and simulating the reverse diffusion process.

The original diffusion model for images \citep{sohl2015deep} operates directly in the image space. That is, $X$ is an image and $Z_1,\dots, Z_T$ are noisy images of the same dimensions as $X$. Therefore, the encoder does not need to be trained because $q(Z_1 \mid X)$ simply adds noise to the image $X$. This makes the model simpler, since only the decoder needs to be learned. However, generating high-quality samples requires dividing both the forward and backward processes into small steps, which can be computationally costly when $T$ is large.

To address this issue, stable diffusion \citep{rombach2022high} uses latent variables $Z_1,\dots, Z_T$ of smaller dimensions, which makes inference faster, but adds the cost of learning an encoder $q(Z_1 \mid X)$ and decoder $p(X\mid Z_1)$. In practice, pre-trained models are often used for $p$ and $q$ to avoid the training cost. Figure~\ref{fig:diffusion-architecture} (top) shows the architecture of a diffusion model for image generation. The standard diffusion model operated of \citet{sohl2015deep} operated in pixel space, and thus did not have the $p$ and $q$ decoder and encoder models. 

Even with the addition of latent variables, at inference time diffusion models still require
many evaluation steps~\citep{yang2023diffusion}. Recent research has focused on reducing the time spent in the inference process by reducing the number of steps required~\citep{song2020score, karras2022elucidating, dockhorn2021score, song2020denoising, lu2022dpm}, or training a better sampler to directly select the best possible steps~\citep{watson2021learning, salimans2022progressive, meng2023distillation}.

\paragraph{Multimodal diffusion models}
In the case of multimodal data, we could also build a multimodal diffusion model as above, say with $X=(X^1,X^2)$ being images and text. However, two challenges emerge. First, the same challenges of building multimodal encoders and decoders as in VAEs. Second, even if we can build models with separate encoders and decoders per modality, the issue is that diffusion models are not as suitable for text generation as they are for image generation. Specifically, while diffusion models for text generation have been developed, e.g., by using a discrete latent space $Z$ with categorical transition probabilities \citep{austin2021structured}, text encoders based on transformers or other sequence-to-sequence models are preferred in practice. As a consequence, multimodal models for both text and images, such as text-to-image generation models, combine text encoders with diffusion models for images. Figure~\ref{fig:diffusion-architecture} (bottom) shows one possible architecture for such a model. As with the other generative approaches, an additional input $Y$ (e.g.,~a text description) can be used to condition the generation.\footnote{Note that, unlike the case of VAEs where the conditioning affects both the encoder $q(Z \mid X, Y)$ and the decoder $p(X \mid Z, Y)$, in the case of diffusion models the conditioning does not affect $q(Z_t \mid Z_{t-1})$ because it a simple noising process.} 
 
Recently, many conditional diffusion models for image generation have been proposed using text and other modalities as the conditioning variables. For example,
\textit{DALL-E}~\citep{ramesh2022hierarchical, betker2023improving} uses the CLIP \citep{radford2021learning} embedding space as a starting point to generate novel images. To this objective, the authors train a decoder to invert the CLIP representation back to images. Working on a space that jointly represents text and images allows one to apply language-guided image manipulations. \citet{betker2023improving} improve the quality of the generated images by performing an automated cleaning and improvement of the training image captions with a dedicated captioning model.
\textit{Stable Diffusion}~\citep{rombach2022high} is able to generate images from an input text. Since directly training in the pixel space is very computationally demanding, the generative part of Stable Diffusion is trained on a lower-dimension feature space, and relies on a UNet~\citep{ronneberger2015u} autoencoder separately pre-trained on a perceptual loss and a patch-based adversarial objective. To condition the generation based on other modality inputs, such as texts or semantic maps, they train a cross-attention layer to project the new modality inputs to the intermediate layers of the UNet.
\textit{Imagen}~\citep{saharia2022photorealistic} train a diffusion model for image generation based on a U-Net image model and a T5 text encoder pre-trained only with text. To condition the image generation based on text, the authors find that using cross-attention significantly outperforms other pooling strategies, and achieves high image-text alignment as well as photo-realistic results.

Other works expanded diffusion models in different directions. For example, \citet{zhang2023adding} increase the controllability of the generated results, \citet{brooks2023instructpix2pix} add instruction-following capabilities for image modification, \citet{ruiz2023dreambooth} improve the consistency of the generated subject's identity by fine-tuning the model with a few images, and \citet{Chen2024mmdiff} propose a multi-modal, multi-task, diffusion model, where multiple input modalities are fused and fed to various decoders to accomplish multiple tasks simultaneously.  

Diffusion models have also been used for sound generation~\citep{yang2023diffsound}, video generation~\citep{jeong2023power, videoworldsimulators2024}, and other modalities~\citep{pmlr-v202-kotelnikov23a, lin2023diffusion}, or multiple modalities simultaneously~\citep{tang2024any, ruan2023mm}. See~\citep{cao2024survey} for a recent survey on diffusion models and its applications.

This rapid progress in diffusion model research shows great potential in their usefulness for recommendation applications. 
\citet{zhu2023tryondiffusion} propose a virtual try-on system based on a diffusion model, which outperforms earlier ones based on GANs. \citet{haokai24mmcdiff} and \citet{jiang2024diffmm} use diffusion models to combine multimodal item information with user-item interaction data.

\subsection{Interactive Multimodal Recommendation Models}
\label{sec:interactive-mm-rec}

As seen in Chapter~4, Large Language Models (LLMs) have been widely used in recommender systems, to do tasks like top-k recommendation, rating prediction or explanation generation~\citep{geng2022recommendation, wu2023survey, he2023large}. In this section we discuss interactive multimodal recommendation models based on LLMs, which have demonstrated impressive generalization capabilities and apparent emergent properties to solve tasks not directly targeted during training~\citep{brown2020language}. 

\begin{figure}[tb]
\centerline{\includegraphics[width=0.7\columnwidth,keepaspectratio]{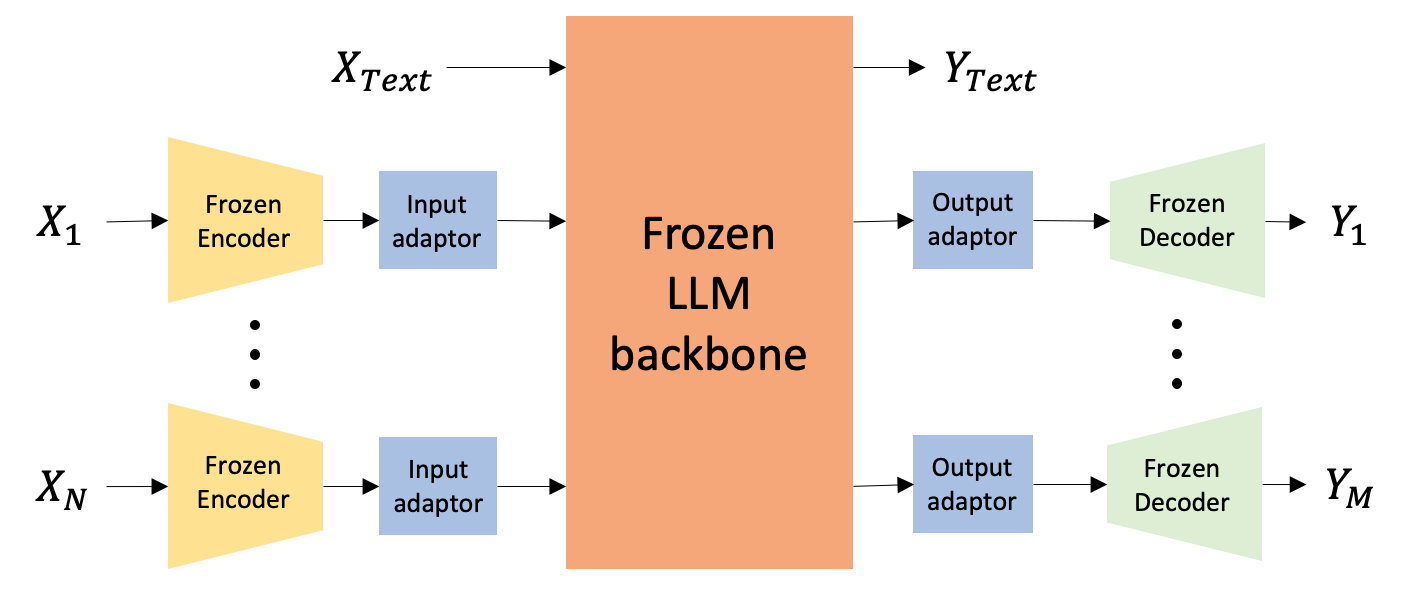}}
\caption{High-level architecture of an MLLM model. 
Each input is processed by a specialized
encoder to obtain modality specific features which are
then projected to a representation adequate for an LLM backbone via suitable adaptors. Similarly, the output of the LLM is then projected to serve as input for specific generators for each modality via a similar adaptor. The input and output modalities are independent.}
\label{fig:mllm-architecture}
\end{figure}

\paragraph{Multimodal Large Language Models} One approach to designing interactive multimodal recommendation systems is to train or adapt specialized \emph{X-to-text} encoders that allow LLMs to accept multimodal input, such as images~\citep{liu2024visual} or other modalities~\citep{wu2023next, tang2023codi2}. These new models are called Multimodal Large Language Models (MLLM), and greatly extend the capabilities of LLMs not only at the input side, by accepting information expressed in different modalities, but also at the output side, where appropriate decoders can be used to allow the model to generate content in various modalities, replacing or complementing the textual answer. Figure~\ref{fig:mllm-architecture} shows a high-level diagram of an MLLM architecture.

With the addition of multiple modalities, MLLMs can become versatile task solvers for recommendation problems. They provide a natural language interface for users to express their queries in multiple modalities, they can tackle complex zero-shot recommendation tasks thanks to their emergent properties, or orchestrate several sub-systems to obtain the best recommendation, they can also generate fluent natural language explanations for a multi-modal recommendation, or even generate documents in different modalities to help the user visualize the products.

As discussed earlier, given the complexity of training large generative models end-to-end, researchers typically assemble systems composed of discriminatively pre-trained components (encoders, decoders and LLM ``reasoning'' models), usually connected by adaptation layers. These adaptation layers are usually pre-trained using paired data for the different modalities, sometimes together with or followed by some form of parameter-efficient fine-tuning of the base models. For example, Low Rank Adaptation (LoRA)~\citep{hu2021lora} freezes the pre-trained parameters and introduces low-rank decomposable trainable adaptation matrices to each transformer layer. This fine-tuning step ensures that representations from different modalities are aligned. In some cases, an existing expert model able to produce textual descriptions of the multi-media content is used in place of adaptation layers~\citep{li2023videochat, yang2024gpt4tools, gao2023llama, wang2024modaverse}. This may lead to lower data needs for adaptation, but could also result in information lost in translation. 

\paragraph{Controller LLMs} Similar to the ``dialogue controller'' described in Chapter~4.7.2, instead of training adaptation layers, another approach is to allow a ``controller'' LLM to use external tools (e.g., foundation models, classical recommendation systems, arbitrary functions), to deal with the multi-modal input and output~\citep{yin2023survey}. This approach has the advantage that it usually involves little or no training. With a carefully constructed prompt explaining all the available tool capabilities and providing usage examples, the controller LLM can create an execution plan with the multi-modal input and generate the desired result in a zero-shot fashion~\citep{zhang2024mm}. Once the plan is complete, the controller LLM will have to re-asses the output, and decide if the desired result has been achieved, or further steps are needed. To improve results, researchers have also tried to use instruction-tuning of the controller LLM to improve the tool selection and planning abilities~\citep{yang2024gpt4tools}. An obvious drawback of this approach is that it requires processing multiple rounds of instructions and multiple rounds of (possibly slow) ``tool'' foundation models to get to the desired result.

\paragraph{Multimodal instruction tuning} As seen in Section~4.7.2, instruction tuning is an important step to make LLMs useful task solvers. It requires the creation of datasets of instruction-formatted examples that will be used as training data for the model. These datasets are usually created by extending input-output pairs from multiple multi-modal datasets, such as COCO Captions \citep{chen2015microsoft}, LAION \citep{schuhmann2022laion} or VQAv2 \citep{balanced_vqa_v2}, with an instruction text defined using prompt templates for the different tasks \citep{dai2024instructblip}. Tasks can be created, for example, by using annotated bounding boxes to define a spatial relationship question, or using an existing image caption as supervision for an image description request.
When no suitable datasets are available, and collecting them from other sources is impractical, 
researchers use generative models to create the examples through self-instruction~\citep{brooks2023instructpix2pix, wang2022self}. Models that extend instructions to multiple modalities have recently been proposed~\citep{li2023instructany2pix, wu2023next}. For example, a user could issue a query such as \textit{``modify [image] to convey the feeling of [music]''}.
With instruction tuning, MLLMs can be used in dialog systems (c.f. Sec~4.7), enriching the conversation with multimodal understanding and generation.

\paragraph{Any-to-text MLLMs}
Recently, many MLLMs that add images to the accepted inputs have been developed: \citet{alayrac2022flamingo} propose \textit{Flamingo}, which use an LLM and a vision encoder trained with a contrastive loss similar to CLIP to build a model able to meet the state of the art performance in various image and video tasks.
\citet{liu2024visual} propose \textit{Llava}, an instruction-tuned multi-modal LLM that is able to accept input in both text and image format, and produce useful textual responses. The authors connect the grid visual embeddings from the last layers of the CLIP image encoder~\citep{radford2021learning} with the Vicuna language decoder~\citep{vicuna2023} using a simple linear adaptation layer, and fine-tunning the model end to end (they keep the visual encoder weights frozen). To train the model, instruction-following training data is generated using GPT-4~\citep{achiam2023gpt} and image-caption datasets, such as COCO~\citep{chen2015microsoft}. \citet{liu2023improved} change the connection layer from a linear projection to a two-layer MLP and obtain better results. \citet{li2023blip} propose \textit{BLIP-2}, introducing a lightweight Query-Transformer (Q-Former), which consists of two transformer modules, to bridge the modality gap between the image encoder output and the LLM, and allow prompts to include both text and image. 

Although image is the modality that received the most attention, some works have addressed adding audio~\citep{deshmukh2023pengi, kong2024audio}, and multiple modalities, like text, image, audio or video~\citep{han2023imagebind, moon2023anymal, lyu2023macaw}. However, these models are still limited to generating only text output.

\paragraph{Any-to-any MLLMs}
As mentioned earlier, to overcome the single-modality output limitation, authors have proposed systems that can both absorb information in multiple modalities, as well as generate response content in different modalities. For example, Next-GPT~\citep{wu2023next} attempts any-to-any modality conversion through an MM-LLM by using state-of-the-art encoders and decoders, connected to the LLM by thin adapter layers. Multi-modality switching instruction tunning is learned using a custom dataset of 5000 high quality samples. After warming up the adaptation layers, the whole system is trained using LoRA with the modality-switching dataset. For input, the authors use ImageBind~\citep{girdhar2023imagebind}, which has been trained to produce aligned representations for image, audio and video, among other modalities, and adapt it using a linear layer to the Vicuna LLM, that does the core reasoning/instruction-following. For the output, small modality-specific transformers are trained to produce the input for three state-of-the-art decoder models, for audio, image and video. \citet{tang2023codi2} use a similar approach, with Llama2~\citep{touvron2023llama} as the core LLM and state-of-the-art diffusion models to generate the multi-modal outputs.

Several companies have released proprietary generalist MLLM-powered chatbots, like OpenAI GPT-4~\citep{achiam2023gpt} and Google Gemini~\citep{team2023gemini} or Anthropic Claude~\citep{anthropic2024claude}. Even though these models are not explicitly trained as recommender systems, they are able to produce a variety of recommendation results, including shopping recommendations. For example, Gemini can receive images, audio and video as input, recognize objects in them, provide general advice for product understanding, and recommend products based on customer input. 

\section{Applications of Multimodal Recommendation Systems}
\label{sec:applications-mm}

Recent developments in multi-modal generative models open the door to many applications in recommender systems. In this section we review some of the most promising directions, in areas including e-commerce, in-context product visualization, marketing, online streaming, and travel and service recommendations.

\paragraph{E-commerce} One of the most direct applications of generative multimodal models for recommendation is e-commerce, where there is a large volume of product and customer data available that can be used to benefit the customer recommendations. Applications range from improving product images~\citep{corneanu2024latentpaint}, names and descriptions~\citep{novgorodov2019generating, shao2021controllable}, to generating reviews~\citet{truong2019multimodal} and review summaries~\citep{amazon2023reviews}, learning to generate better recommendation~\citep{xiao2022abstract, liu2024rec, karra2024interarec}, and answering user questions~\citep{deng2022toward}. 

\citet{karra2024interarec} propose to use multimodal large language models to improve recommendations by better understanding the behavior of users in e-commerce websites. As the user navigates during a browsing session, high-frequency screenshots are captured and provided to an MLLM together with specific prompts requesting to extract information such as price ranges, product categories and brand preferences, to generate a user behavioral summary. Next, this summary is provided to an LLM with tool-using abilities to derive features and constraints from the input, and use a recommender system to generate the final recommendation. \citet{liu2024rec} describe the limitations of current MLLM when used with multiple images as input in the prompt. To improve the performance, they propose to process the list of products interacted by the user as pairs of image and title to obtain text descriptions of the products. These descriptions can then be used in lieu of the images when using the interaction history as in-context-learning to generate new recommendations for a user with an MLLM.
\citet{truong2019multimodal} propose a system to generate multi-modal reviews. The proposed system uses item and user embeddings, obtained via matrix factorization, to predict the rating and compose a review text with a Long Short-Term Memory network. If a review image is available, it is also used to condition the text generation.

\paragraph{In-context product visualization} Applications such as ``virtual try on'' or ``view in your room'' augment an image or video with products such as clothes~\citep{yuan2013mixed, han2018viton}, sunglasses, or even makeup~\citep{borges2019virtual, beauty2023compare} to help users visualize how they would look in themselves, or how furniture or appliances would look in the context of their home~\citep{reuksupasompon2018ar, amazonARtech, amazonARcsa}, before making a purchase decision. A traditional approach for these tasks is Augmented Reality (AR), that mixes real images, obtained from a camera feed, with virtual objects to generate novel views in real time. While AR has been used in numerous applications, ranging from education~\citep{billinghurst2002augmented} to assisting surgeons in medical operations~\citep{dennler2021augmented}, recent diffusion-based image generation models can be used to further improve virtual-try-on experiences~\citep{wang2024stablegarment, xu2024ootdiffusion, wang2024texfit}, or generate outfits to try out~\citep{xu2024difashion}, and make them more controllable~\citep{seyfioglu2024diffuse}.

\paragraph{Marketing} In marketing, multimodal generative models can be used to create personalized advertisement images and videos from product imagery and customer preferences to increase the probability of engagement~\citep{wang2023generate, chen2021automated}. \citet{wei2022towards} generates personalized bundles and creates a customized image for display. \citet{shilova2023adbooster} fine-tune a Stable Diffusion model to generate personalized images by outpainting input images without modifying the targeted object. For training, they leverage a U$^2$-Net segmentation network, and a BLIP model to generate masks and captions for a collection of training images, that the model will then learn to reconstruct. With adequate guardrails in place, generative models could also be used to synthesize personalized multimodal ad content like text~\citep{loukili23marketing}, images~\citep{mayahi2022impact} or video~\citep{liu2023ai}.

\paragraph{Streaming services} Online video and audio streaming services strive to recommend the most valuable multimedia content to each user in order to maximize usage, ad revenue or click-through rate. Long and short-form video, music, audiobooks, podcasts and radio have different recommendation requirements, but the very content to recommend comes in multiple modalities that can be used to improve the suggestions. Even though most works on streaming content recommendation rely on user behavior and content metadata, recent works have applied multimodal learning to audio~\citep{jones2023learning, chen2021learning, huang2020large, deldjoo2024content} and video~\citep{lei20mmmicrov, wei2019mmgcn, yi2022multi, sun2022long} recommendation. Due to their extensive pre-training, large multi-modal generative models can further enhance the user experience in a content streaming recommendation setting by blending content understanding with personalization and generation, allowing them to complete tasks like answering to fine-grained content-related questions in natural language (e.g., ``Does this movie contain a car chase scene that I will like?''), or generating personalized content to fulfill a user request (e.g., audio and music generation~\citep{briot2017deep, lam2024efficient, vyas2023audiobox, dhariwal2020jukebox}). 

\paragraph{Travel and service recommendations} Services ranging from theme parks and concert venues to restaurants, auto mechanics, and laundry services receive customer ratings, reviews and clicks in many online platforms. Better understanding of contextual details such as location characteristics, services offered, past user experiences and popularity factors through multi-modal information, could lead to better and more personalized recommendations. Furthermore, future work could tackle comprehensive products such as interactive recommendation systems able to help users through the whole process of planning and booking complex multi-faceted events such as weddings (e.g.,~venue, menu, decoration, music) or travel (e.g.,~destination, transport, hotel, restaurant, activities, practical tips) in a conversational way~\citep{xie2024travelplanner}, accepting and incorporating user feedback and explaining the recommendations.
\citet{yan2023personalized} generates explanations for recommendations focusing on making them informative and diverse. For that, they start by selecting a set of images for a given user and business using a detrimental point process that leverages CLIP features from the user history, and the business images. Then, they use a GPT-2-powered multi-modal decoder, trained with a personalized cross-modal contrastive loss, to generate natural language explanations. The results show that the proposed method produces more informative and diverse explanations compared to text-only alternatives.

\backmatter

\printbibliography

@inproceedings{Kingma2014,
  author = {Kingma, Diederik P. and Welling, Max},
  booktitle = {Proceedings of the 2nd International Conference on Learning Representations, {ICLR} 2014},
  title = {{Auto-Encoding Variational Bayes}},
  year = 2014
}

@article{wu2018multimodal,
  title={Multimodal generative models for scalable weakly-supervised learning},
  author={Wu, Mike and Goodman, Noah},
  journal={Advances in neural information processing systems},
  volume={31},
  year={2018}
}

@article{shi2019variational,
  title={Variational mixture-of-experts autoencoders for multi-modal deep generative models},
  author={Shi, Yuge and Paige, Brooks and Torr, Philip and others},
  journal={Advances in neural information processing systems},
  volume={32},
  year={2019}
}

@article{sutter2020multimodal,
  title={Multimodal generative learning utilizing jensen-shannon-divergence},
  author={Sutter, Thomas and Daunhawer, Imant and Vogt, Julia},
  journal={Advances in neural information processing systems},
  volume={33},
  pages={6100--6110},
  year={2020}
}

@article{yi2021cross,
  title={Cross-modal variational auto-encoder for content-based micro-video background music recommendation},
  author={Yi, Jing and Zhu, Yaochen and Xie, Jiayi and Chen, Zhenzhong},
  journal={IEEE Transactions on Multimedia},
  volume={25},
  pages={515--528},
  year={2021},
  publisher={IEEE}
}

@article{oord2018representation,
  title={Representation learning with contrastive predictive coding},
  author={Oord, Aaron van den and Li, Yazhe and Vinyals, Oriol},
  journal={arXiv preprint arXiv:1807.03748},
  year={2018}
}

@inproceedings{karamanolakis2018item,
  title={Item recommendation with variational autoencoders and heterogeneous priors},
  author={Karamanolakis, Giannis and Cherian, Kevin Raji and Narayan, Ananth Ravi and Yuan, Jie and Tang, Da and Jebara, Tony},
  booktitle={Proceedings of the 3rd Workshop on Deep Learning for Recommender Systems},
  pages={10--14},
  year={2018}
}

@article{he2023large,
  title={Large language models as zero-shot conversational recommenders},
  author={He, Zhankui and Xie, Zhouhang and Jha, Rahul and Steck, Harald and Liang, Dawen and Feng, Yesu and Majumder, Bodhisattwa Prasad and Kallus, Nathan and McAuley, Julian},
  journal={arXiv preprint arXiv:2308.10053},
  year={2023}
}

@inproceedings{geng2022recommendation,
  title={Recommendation as language processing (rlp): A unified pretrain, personalized prompt \& predict paradigm (p5)},
  author={Geng, Shijie and Liu, Shuchang and Fu, Zuohui and Ge, Yingqiang and Zhang, Yongfeng},
  booktitle={Proceedings of the 16th ACM Conference on Recommender Systems},
  pages={299--315},
  year={2022}
}

@inproceedings{wei2023multi,
  title={Multi-Modal Self-Supervised Learning for Recommendation},
  author={Wei, Wei and Huang, Chao and Xia, Lianghao and Zhang, Chuxu},
  booktitle={Proceedings of the ACM Web Conference 2023},
  pages={790--800},
  year={2023}
}

@inproceedings{radford2021learning,
  title={Learning transferable visual models from natural language supervision},
  author={Radford, Alec and Kim, Jong Wook and Hallacy, Chris and Ramesh, Aditya and Goh, Gabriel and Agarwal, Sandhini and Sastry, Girish and Askell, Amanda and Mishkin, Pamela and Clark, Jack and others},
  booktitle={International conference on machine learning},
  pages={8748--8763},
  year={2021},
  organization={PMLR}
}

@article{li2021align,
  title={Align before fuse: Vision and language representation learning with momentum distillation},
  author={Li, Junnan and Selvaraju, Ramprasaath and Gotmare, Akhilesh and Joty, Shafiq and Xiong, Caiming and Hoi, Steven Chu Hong},
  journal={Advances in neural information processing systems},
  volume={34},
  pages={9694--9705},
  year={2021}
}

@inproceedings{goodfellow2014generative,
  title={Generative adversarial nets},
  author={Goodfellow, Ian and Pouget-Abadie, Jean and Mirza, Mehdi and Xu, Bing and Warde-Farley, David and Ozair, Sherjil and Courville, Aaron and Bengio, Yoshua},
  booktitle={Advances in Neural Information Processing Systems},
  year={2014}
}

@article{anderson1982reverse,
  title={Reverse-time diffusion equation models},
  author={Anderson, Brian DO},
  journal={Stochastic Processes and their Applications},
  volume={12},
  number={3},
  pages={313--326},
  year={1982},
  publisher={Elsevier}
}

@article{austin2021structured,
  author       = {Jacob Austin and
                  Daniel D. Johnson and
                  Jonathan Ho and
                  Daniel Tarlow and
                  Rianne van den Berg},
  title        = {Structured Denoising Diffusion Models in Discrete State-Spaces},
  journal      = {CoRR},
  volume       = {abs/2107.03006},
  year         = {2021},
  url          = {https://arxiv.org/abs/2107.03006},
  eprinttype    = {arXiv},
  eprint       = {2107.03006},
  bibsource    = {dblp computer science bibliography, https://dblp.org}
}

@inproceedings{Feller1949,
  title={On the theory of stochastic processes, with particular reference to applications},
  author={Feller, W},
  booktitle={Berkeley Symposium on Mathematical Statistics and Probability},
  pages={403–432},
  year={1949}
}

@article{guo2019deep,
  title={Deep multimodal representation learning: A survey},
  author={Guo, Wenzhong and Wang, Jianwen and Wang, Shiping},
  journal={Ieee Access},
  volume={7},
  pages={63373--63394},
  year={2019},
  publisher={IEEE}
}

@inproceedings{10.5555/3495724.3497510,
author = {Grill, Jean-Bastien and Strub, Florian and Altch\'{e}, Florent and Tallec, Corentin and Richemond, Pierre H. and Buchatskaya, Elena and Doersch, Carl and Pires, Bernardo Avila and Guo, Zhaohan Daniel and Azar, Mohammad Gheshlaghi and Piot, Bilal and Kavukcuoglu, Koray and Munos, R\'{e}mi and Valko, Michal},
title = {Bootstrap your own latent a new approach to self-supervised learning},
year = {2020},
isbn = {9781713829546},
publisher = {Curran Associates Inc.},
address = {Red Hook, NY, USA},
booktitle = {Proceedings of the 34th International Conference on Neural Information Processing Systems},
articleno = {1786},
numpages = {14},
location = {Vancouver, BC, Canada},
series = {NIPS'20}
}

@InProceedings{Caron_2021_ICCV,
    author    = {Caron, Mathilde and Touvron, Hugo and Misra, Ishan and J\'egou, Herv\'e and Mairal, Julien and Bojanowski, Piotr and Joulin, Armand},
    title     = {Emerging Properties in Self-Supervised Vision Transformers},
    booktitle = {Proceedings of the IEEE/CVF International Conference on Computer Vision (ICCV)},
    month     = {October},
    year      = {2021},
    pages     = {9650-9660}
}

@inproceedings{saeed2021contrastive,
  title={Contrastive learning of general-purpose audio representations},
  author={Saeed, Aaqib and Grangier, David and Zeghidour, Neil},
  booktitle={ICASSP 2021-2021 IEEE International Conference on Acoustics, Speech and Signal Processing (ICASSP)},
  pages={3875--3879},
  year={2021},
  organization={IEEE}
}

@inproceedings{cheng2020look,
  title={Look, listen, and attend: Co-attention network for self-supervised audio-visual representation learning},
  author={Cheng, Ying and Wang, Ruize and Pan, Zhihao and Feng, Rui and Zhang, Yuejie},
  booktitle={Proceedings of the 28th ACM International Conference on Multimedia},
  pages={3884--3892},
  year={2020}
}

@inproceedings{won2020data,
  title={Data-driven harmonic filters for audio representation learning},
  author={Won, Minz and Chun, Sanghyuk and Nieto, Oriol and Serrc, Xavier},
  booktitle={ICASSP 2020-2020 IEEE International Conference on Acoustics, Speech and Signal Processing (ICASSP)},
  pages={536--540},
  year={2020},
  organization={IEEE}
}

@inproceedings{wang2022towards,
  title={Towards learning universal audio representations},
  author={Wang, Luyu and Luc, Pauline and Wu, Yan and Recasens, Adria and Smaira, Lucas and Brock, Andrew and Jaegle, Andrew and Alayrac, Jean-Baptiste and Dieleman, Sander and Carreira, Joao and others},
  booktitle={ICASSP 2022-2022 IEEE International Conference on Acoustics, Speech and Signal Processing (ICASSP)},
  pages={4593--4597},
  year={2022},
  organization={IEEE}
}

@inproceedings{he2020momentum,
  title={Momentum contrast for unsupervised visual representation learning},
  author={He, Kaiming and Fan, Haoqi and Wu, Yuxin and Xie, Saining and Girshick, Ross},
  booktitle={Proceedings of the IEEE/CVF conference on computer vision and pattern recognition},
  pages={9729--9738},
  year={2020}
}

@inproceedings{gutmann2010noise,
  title={Noise-contrastive estimation: A new estimation principle for unnormalized statistical models},
  author={Gutmann, Michael and Hyv{\"a}rinen, Aapo},
  booktitle={Proceedings of the thirteenth international conference on artificial intelligence and statistics},
  pages={297--304},
  year={2010},
  organization={JMLR Workshop and Conference Proceedings}
}

@inproceedings{hager2023best,
  title={Best of Both Worlds: Multimodal Contrastive Learning with Tabular and Imaging Data},
  author={Hager, Paul and Menten, Martin J and Rueckert, Daniel},
  booktitle={Proceedings of the IEEE/CVF Conference on Computer Vision and Pattern Recognition},
  pages={23924--23935},
  year={2023}
}

@inproceedings{chen2020simple,
  title={A simple framework for contrastive learning of visual representations},
  author={Chen, Ting and Kornblith, Simon and Norouzi, Mohammad and Hinton, Geoffrey},
  booktitle={International conference on machine learning},
  pages={1597--1607},
  year={2020},
  organization={PMLR}
}

@inproceedings{huang2023multimodal,
  title={Multimodal Contrastive Learning and Tabular Attention for Automated Alzheimer's Disease Prediction},
  author={Huang, Weichen},
  booktitle={Proceedings of the IEEE/CVF International Conference on Computer Vision},
  pages={2473--2482},
  year={2023}
}

@inproceedings{chen2020generative,
  title={Generative pretraining from pixels},
  author={Chen, Mark and Radford, Alec and Child, Rewon and Wu, Jeffrey and Jun, Heewoo and Luan, David and Sutskever, Ilya},
  booktitle={International conference on machine learning},
  pages={1691--1703},
  year={2020},
  organization={PMLR}
}

@inproceedings{girdhar2023imagebind,
  title={Imagebind: One embedding space to bind them all},
  author={Girdhar, Rohit and El-Nouby, Alaaeldin and Liu, Zhuang and Singh, Mannat and Alwala, Kalyan Vasudev and Joulin, Armand and Misra, Ishan},
  booktitle={Proceedings of the IEEE/CVF Conference on Computer Vision and Pattern Recognition},
  pages={15180--15190},
  year={2023}
}

@article{pham2023combined,
  title={Combined scaling for zero-shot transfer learning},
  author={Pham, Hieu and Dai, Zihang and Ghiasi, Golnaz and Kawaguchi, Kenji and Liu, Hanxiao and Yu, Adams Wei and Yu, Jiahui and Chen, Yi-Ting and Luong, Minh-Thang and Wu, Yonghui and others},
  journal={Neurocomputing},
  volume={555},
  pages={126658},
  year={2023},
  publisher={Elsevier}
}

@inproceedings{zhai2022lit,
  title={Lit: Zero-shot transfer with locked-image text tuning},
  author={Zhai, Xiaohua and Wang, Xiao and Mustafa, Basil and Steiner, Andreas and Keysers, Daniel and Kolesnikov, Alexander and Beyer, Lucas},
  booktitle={Proceedings of the IEEE/CVF Conference on Computer Vision and Pattern Recognition},
  pages={18123--18133},
  year={2022}
}

@inproceedings{cherti2023reproducible,
  title={Reproducible scaling laws for contrastive language-image learning},
  author={Cherti, Mehdi and Beaumont, Romain and Wightman, Ross and Wortsman, Mitchell and Ilharco, Gabriel and Gordon, Cade and Schuhmann, Christoph and Schmidt, Ludwig and Jitsev, Jenia},
  booktitle={Proceedings of the IEEE/CVF Conference on Computer Vision and Pattern Recognition},
  pages={2818--2829},
  year={2023}
}

@article{cao2023less,
  title={Less is More: Removing Text-regions Improves CLIP Training Efficiency and Robustness},
  author={Cao, Liangliang and Zhang, Bowen and Chen, Chen and Yang, Yinfei and Du, Xianzhi and Zhang, Wencong and Lu, Zhiyun and Zheng, Yantao},
  journal={arXiv preprint arXiv:2305.05095},
  year={2023}
}

@inproceedings{he2022masked,
  title={Masked autoencoders are scalable vision learners},
  author={He, Kaiming and Chen, Xinlei and Xie, Saining and Li, Yanghao and Doll{\'a}r, Piotr and Girshick, Ross},
  booktitle={Proceedings of the IEEE/CVF conference on computer vision and pattern recognition},
  pages={16000--16009},
  year={2022}
}

@article{li2023multimodal,
  title={Multimodal foundation models: From specialists to general-purpose assistants},
  author={Li, Chunyuan and Gan, Zhe and Yang, Zhengyuan and Yang, Jianwei and Li, Linjie and Wang, Lijuan and Gao, Jianfeng},
  journal={arXiv preprint arXiv:2309.10020},
  volume={1},
  number={2},
  pages={2},
  year={2023}
}

@article{shen2022k,
  title={K-lite: Learning transferable visual models with external knowledge},
  author={Shen, Sheng and Li, Chunyuan and Hu, Xiaowei and Xie, Yujia and Yang, Jianwei and Zhang, Pengchuan and Gan, Zhe and Wang, Lijuan and Yuan, Lu and Liu, Ce and others},
  journal={Advances in Neural Information Processing Systems},
  volume={35},
  pages={15558--15573},
  year={2022}
}

@article{fan2023improving,
  title={Improving CLIP Training with Language Rewrites},
  author={Fan, Lijie and Krishnan, Dilip and Isola, Phillip and Katabi, Dina and Tian, Yonglong},
  journal={arXiv preprint arXiv:2305.20088},
  year={2023}
}

@article{chen2023stair,
  title={STAIR: Learning Sparse Text and Image Representation in Grounded Tokens},
  author={Chen, Chen and Zhang, Bowen and Cao, Liangliang and Shen, Jiguang and Gunter, Tom and Jose, Albin Madappally and Toshev, Alexander and Shlens, Jonathon and Pang, Ruoming and Yang, Yinfei},
  journal={arXiv preprint arXiv:2301.13081},
  year={2023}
}

@inproceedings{carlsson2022cross,
  title={Cross-lingual and multilingual clip},
  author={Carlsson, Fredrik and Eisen, Philipp and Rekathati, Faton and Sahlgren, Magnus},
  booktitle={Proceedings of the Thirteenth Language Resources and Evaluation Conference},
  pages={6848--6854},
  year={2022}
}

@inproceedings{dehdashtianfairerclip,
  title={FAIRERCLIP: DEBIASING ZERO-SHOT PREDICTIONS OF CLIP IN RKHSS},
  author={Dehdashtian, Sepehr and Wang, Lan and Boddeti, Vishnu Naresh},
  booktitle={ICLR2024? VERIFY},
  pages={},
  year={2024}
}

@inproceedings{huang2024froster,
title={{FROSTER}: Frozen {CLIP} is A Strong Teacher for Open-Vocabulary Action Recognition},
author={Xiaohu Huang and Hao Zhou and Kun Yao and Kai Han},
booktitle={The Twelfth International Conference on Learning Representations},
year={2024}
}

@inproceedings{jia2021scaling,
  title={Scaling up visual and vision-language representation learning with noisy text supervision},
  author={Jia, Chao and Yang, Yinfei and Xia, Ye and Chen, Yi-Ting and Parekh, Zarana and Pham, Hieu and Le, Quoc and Sung, Yun-Hsuan and Li, Zhen and Duerig, Tom},
  booktitle={International conference on machine learning},
  pages={4904--4916},
  year={2021},
  organization={PMLR}
}

@inproceedings{li2022blip,
  title={Blip: Bootstrapping language-image pre-training for unified vision-language understanding and generation},
  author={Li, Junnan and Li, Dongxu and Xiong, Caiming and Hoi, Steven},
  booktitle={International Conference on Machine Learning},
  pages={12888--12900},
  year={2022},
  organization={PMLR}
}

@inproceedings{rombach2022high,
  title={High-resolution image synthesis with latent diffusion models},
  author={Rombach, Robin and Blattmann, Andreas and Lorenz, Dominik and Esser, Patrick and Ommer, Bj{\"o}rn},
  booktitle={Proceedings of the IEEE/CVF conference on computer vision and pattern recognition},
  pages={10684--10695},
  year={2022}
}

@article{ramesh2022hierarchical,
  title={Hierarchical text-conditional image generation with clip latents},
  author={Ramesh, Aditya and Dhariwal, Prafulla and Nichol, Alex and Chu, Casey and Chen, Mark},
  journal={arXiv preprint arXiv:2204.06125},
  volume={1},
  number={2},
  pages={3},
  year={2022}
}

@article{yang2023diffsound,
  title={Diffsound: Discrete diffusion model for text-to-sound generation},
  author={Yang, Dongchao and Yu, Jianwei and Wang, Helin and Wang, Wen and Weng, Chao and Zou, Yuexian and Yu, Dong},
  journal={IEEE/ACM Transactions on Audio, Speech, and Language Processing},
  year={2023},
  publisher={IEEE}
}

@inproceedings{jeong2023power,
  title={The power of sound (tpos): Audio reactive video generation with stable diffusion},
  author={Jeong, Yujin and Ryoo, Wonjeong and Lee, Seunghyun and Seo, Dabin and Byeon, Wonmin and Kim, Sangpil and Kim, Jinkyu},
  booktitle={Proceedings of the IEEE/CVF International Conference on Computer Vision},
  pages={7822--7832},
  year={2023}
}

@article{tang2024any,
  title={Any-to-any generation via composable diffusion},
  author={Tang, Zineng and Yang, Ziyi and Zhu, Chenguang and Zeng, Michael and Bansal, Mohit},
  journal={Advances in Neural Information Processing Systems},
  volume={36},
  year={2024}
}

@article{cao2024survey,
  title={A survey on generative diffusion models},
  author={Cao, Hanqun and Tan, Cheng and Gao, Zhangyang and Xu, Yilun and Chen, Guangyong and Heng, Pheng-Ann and Li, Stan Z},
  journal={IEEE Transactions on Knowledge and Data Engineering},
  year={2024},
  publisher={IEEE}
}

@article{rahate2022multimodal,
  title={Multimodal co-learning: Challenges, applications with datasets, recent advances and future directions},
  author={Rahate, Anil and Walambe, Rahee and Ramanna, Sheela and Kotecha, Ketan},
  journal={Information Fusion},
  volume={81},
  pages={203--239},
  year={2022},
  publisher={Elsevier}
}

@inproceedings{sohl2015deep,
  title={Deep unsupervised learning using nonequilibrium thermodynamics},
  author={Sohl-Dickstein, Jascha and Weiss, Eric and Maheswaranathan, Niru and Ganguli, Surya},
  booktitle={International conference on machine learning},
  pages={2256--2265},
  year={2015},
  organization={PMLR}
}

@article{saharia2022photorealistic,
  title={Photorealistic text-to-image diffusion models with deep language understanding},
  author={Saharia, Chitwan and Chan, William and Saxena, Saurabh and Li, Lala and Whang, Jay and Denton, Emily L and Ghasemipour, Kamyar and Gontijo Lopes, Raphael and Karagol Ayan, Burcu and Salimans, Tim and others},
  journal={Advances in Neural Information Processing Systems},
  volume={35},
  pages={36479--36494},
  year={2022}
}

@article{yang2023diffusion,
  title={Diffusion models: A comprehensive survey of methods and applications},
  author={Yang, Ling and Zhang, Zhilong and Song, Yang and Hong, Shenda and Xu, Runsheng and Zhao, Yue and Zhang, Wentao and Cui, Bin and Yang, Ming-Hsuan},
  journal={ACM Computing Surveys},
  volume={56},
  number={4},
  pages={1--39},
  year={2023},
  publisher={ACM New York, NY, USA}
}

@article{song2020score,
  title={Score-based generative modeling through stochastic differential equations},
  author={Song, Yang and Sohl-Dickstein, Jascha and Kingma, Diederik P and Kumar, Abhishek and Ermon, Stefano and Poole, Ben},
  journal={arXiv preprint arXiv:2011.13456},
  year={2020}
}

@article{karras2022elucidating,
  title={Elucidating the design space of diffusion-based generative models},
  author={Karras, Tero and Aittala, Miika and Aila, Timo and Laine, Samuli},
  journal={Advances in Neural Information Processing Systems},
  volume={35},
  pages={26565--26577},
  year={2022}
}

@article{dockhorn2021score,
  title={Score-based generative modeling with critically-damped langevin diffusion},
  author={Dockhorn, Tim and Vahdat, Arash and Kreis, Karsten},
  journal={arXiv preprint arXiv:2112.07068},
  year={2021}
}

@article{song2020denoising,
  title={Denoising diffusion implicit models},
  author={Song, Jiaming and Meng, Chenlin and Ermon, Stefano},
  journal={arXiv preprint arXiv:2010.02502},
  year={2020}
}

@article{lu2022dpm,
  title={Dpm-solver: A fast ode solver for diffusion probabilistic model sampling in around 10 steps},
  author={Lu, Cheng and Zhou, Yuhao and Bao, Fan and Chen, Jianfei and Li, Chongxuan and Zhu, Jun},
  journal={Advances in Neural Information Processing Systems},
  volume={35},
  pages={5775--5787},
  year={2022}
}

@inproceedings{watson2021learning,
  title={Learning fast samplers for diffusion models by differentiating through sample quality},
  author={Watson, Daniel and Chan, William and Ho, Jonathan and Norouzi, Mohammad},
  booktitle={International Conference on Learning Representations},
  year={2021}
}

@article{salimans2022progressive,
  title={Progressive distillation for fast sampling of diffusion models},
  author={Salimans, Tim and Ho, Jonathan},
  journal={arXiv preprint arXiv:2202.00512},
  year={2022}
}

@inproceedings{meng2023distillation,
  title={On distillation of guided diffusion models},
  author={Meng, Chenlin and Rombach, Robin and Gao, Ruiqi and Kingma, Diederik and Ermon, Stefano and Ho, Jonathan and Salimans, Tim},
  booktitle={Proceedings of the IEEE/CVF Conference on Computer Vision and Pattern Recognition},
  pages={14297--14306},
  year={2023}
}

@article{betker2023improving,
  title={Improving image generation with better captions},
  author={Betker, James and Goh, Gabriel and Jing, Li and Brooks, Tim and Wang, Jianfeng and Li, Linjie and Ouyang, Long and Zhuang, Juntang and Lee, Joyce and Guo, Yufei and others},
  journal={Computer Science. https://cdn. openai. com/papers/dall-e-3. pdf},
  volume={2},
  number={3},
  pages={8},
  year={2023}
}

@InProceedings{pmlr-v202-kotelnikov23a,
  title = 	 {{T}ab{DDPM}: Modelling Tabular Data with Diffusion Models},
  author =       {Kotelnikov, Akim and Baranchuk, Dmitry and Rubachev, Ivan and Babenko, Artem},
  booktitle = 	 {Proceedings of the 40th International Conference on Machine Learning},
  pages = 	 {17564--17579},
  year = 	 {2023},
  editor = 	 {Krause, Andreas and Brunskill, Emma and Cho, Kyunghyun and Engelhardt, Barbara and Sabato, Sivan and Scarlett, Jonathan},
  volume = 	 {202},
  series = 	 {Proceedings of Machine Learning Research},
  month = 	 {23--29 Jul},
  publisher =    {PMLR},
  url = 	 {https://proceedings.mlr.press/v202/kotelnikov23a.html}
}

@article{li2023videochat,
  title={Videochat: Chat-centric video understanding},
  author={Li, KunChang and He, Yinan and Wang, Yi and Li, Yizhuo and Wang, Wenhai and Luo, Ping and Wang, Yali and Wang, Limin and Qiao, Yu},
  journal={arXiv preprint arXiv:2305.06355},
  year={2023}
}

@article{yang2024gpt4tools,
  title={Gpt4tools: Teaching large language model to use tools via self-instruction},
  author={Yang, Rui and Song, Lin and Li, Yanwei and Zhao, Sijie and Ge, Yixiao and Li, Xiu and Shan, Ying},
  journal={Advances in Neural Information Processing Systems},
  volume={36},
  year={2024}
}

@article{gao2023llama,
  title={Llama-adapter v2: Parameter-efficient visual instruction model},
  author={Gao, Peng and Han, Jiaming and Zhang, Renrui and Lin, Ziyi and Geng, Shijie and Zhou, Aojun and Zhang, Wei and Lu, Pan and He, Conghui and Yue, Xiangyu and others},
  journal={arXiv preprint arXiv:2304.15010},
  year={2023}
}

@article{wang2024modaverse,
  title={ModaVerse: Efficiently Transforming Modalities with LLMs},
  author={Wang, Xinyu and Zhuang, Bohan and Wu, Qi},
  journal={arXiv preprint arXiv:2401.06395},
  year={2024}
}

@article{yin2023survey,
  title={A survey on multimodal large language models},
  author={Yin, Shukang and Fu, Chaoyou and Zhao, Sirui and Li, Ke and Sun, Xing and Xu, Tong and Chen, Enhong},
  journal={arXiv preprint arXiv:2306.13549},
  year={2023}
}

@article{zhang2024mm,
  title={Mm-llms: Recent advances in multimodal large language models},
  author={Zhang, Duzhen and Yu, Yahan and Li, Chenxing and Dong, Jiahua and Su, Dan and Chu, Chenhui and Yu, Dong},
  journal={arXiv preprint arXiv:2401.13601},
  year={2024}
}

@article{dai2024instructblip,
  title={Instructblip: Towards general-purpose vision-language models with instruction tuning},
  author={Dai, Wenliang and Li, Junnan and Li, Dongxu and Tiong, Anthony Meng Huat and Zhao, Junqi and Wang, Weisheng and Li, Boyang and Fung, Pascale N and Hoi, Steven},
  journal={Advances in Neural Information Processing Systems},
  volume={36},
  year={2024}
}

@inproceedings{brooks2023instructpix2pix,
  title={Instructpix2pix: Learning to follow image editing instructions},
  author={Brooks, Tim and Holynski, Aleksander and Efros, Alexei A},
  booktitle={Proceedings of the IEEE/CVF Conference on Computer Vision and Pattern Recognition},
  pages={18392--18402},
  year={2023}
}

@article{wang2022self,
  title={Self-instruct: Aligning language models with self-generated instructions},
  author={Wang, Yizhong and Kordi, Yeganeh and Mishra, Swaroop and Liu, Alisa and Smith, Noah A and Khashabi, Daniel and Hajishirzi, Hannaneh},
  journal={arXiv preprint arXiv:2212.10560},
  year={2022}
}

@article{li2023instructany2pix,
  title={InstructAny2Pix: Flexible Visual Editing via Multimodal Instruction Following},
  author={Li, Shufan and Singh, Harkanwar and Grover, Aditya},
  journal={arXiv preprint arXiv:2312.06738},
  year={2023}
}

@article{wu2023next,
  title={Next-gpt: Any-to-any multimodal llm},
  author={Wu, Shengqiong and Fei, Hao and Qu, Leigang and Ji, Wei and Chua, Tat-Seng},
  journal={arXiv preprint arXiv:2309.05519},
  year={2023}
}

@article{liu2024visual,
  title={Visual instruction tuning},
  author={Liu, Haotian and Li, Chunyuan and Wu, Qingyang and Lee, Yong Jae},
  journal={Advances in neural information processing systems},
  volume={36},
  year={2024}
}

@article{achiam2023gpt,
  title={Gpt-4 technical report},
  author={Achiam, Josh and Adler, Steven and Agarwal, Sandhini and Ahmad, Lama and Akkaya, Ilge and Aleman, Florencia Leoni and Almeida, Diogo and Altenschmidt, Janko and Altman, Sam and Anadkat, Shyamal and others},
  journal={arXiv preprint arXiv:2303.08774},
  year={2023}
}

@article{liu2023improved,
  title={Improved baselines with visual instruction tuning},
  author={Liu, Haotian and Li, Chunyuan and Li, Yuheng and Lee, Yong Jae},
  journal={arXiv preprint arXiv:2310.03744},
  year={2023}
}

@article{lyu2023macaw,
  title={Macaw-llm: Multi-modal language modeling with image, audio, video, and text integration},
  author={Lyu, Chenyang and Wu, Minghao and Wang, Longyue and Huang, Xinting and Liu, Bingshuai and Du, Zefeng and Shi, Shuming and Tu, Zhaopeng},
  journal={arXiv preprint arXiv:2306.09093},
  year={2023}
}

@article{tang2023codi2,
	title={CoDi-2: In-Context, Interleaved, and Interactive Any-to-Any Generation}, 
	author={Zineng Tang and Ziyi Yang and Mahmoud Khademi and Yang Liu and Chenguang Zhu and Mohit Bansal},
	year={2023},
	eprint={2311.18775},
	archivePrefix={arXiv},
	primaryClass={cs.CV}
}

@article{touvron2023llama,
  title={Llama 2: Open foundation and fine-tuned chat models},
  author={Touvron, Hugo and Martin, Louis and Stone, Kevin and Albert, Peter and Almahairi, Amjad and Babaei, Yasmine and Bashlykov, Nikolay and Batra, Soumya and Bhargava, Prajjwal and Bhosale, Shruti and others},
  journal={arXiv preprint arXiv:2307.09288},
  year={2023}
}

@misc{vicuna2023,
    title = {Vicuna: An Open-Source Chatbot Impressing GPT-4 with 90\%* ChatGPT Quality},
    url = {https://lmsys.org/blog/2023-03-30-vicuna/},
    author = {Chiang, Wei-Lin and Li, Zhuohan and Lin, Zi and Sheng, Ying and Wu, Zhanghao and Zhang, Hao and Zheng, Lianmin and Zhuang, Siyuan and Zhuang, Yonghao and Gonzalez, Joseph E. and Stoica, Ion and Xing, Eric P.},
    month = {March},
    year = {2023}
}

@article{rosenbaum2022clasp,
  title={CLASP: Few-Shot Cross-Lingual Data Augmentation for Semantic Parsing},
  author={Rosenbaum, Andy and Soltan, Saleh and Hamza, Wael and Saffari, Amir and Damonte, Marco and Groves, Isabel},
  journal={AACL-IJCNLP 2022},
  pages={444},
  year={2022}
}

@article{nguyen2024dataset,
  title={Dataset diffusion: Diffusion-based synthetic data generation for pixel-level semantic segmentation},
  author={Nguyen, Quang and Vu, Truong and Tran, Anh and Nguyen, Khoi},
  journal={Advances in Neural Information Processing Systems},
  volume={36},
  year={2024}
}

@inproceedings{kirillov2023segment,
  title={Segment anything},
  author={Kirillov, Alexander and Mintun, Eric and Ravi, Nikhila and Mao, Hanzi and Rolland, Chloe and Gustafson, Laura and Xiao, Tete and Whitehead, Spencer and Berg, Alexander C and Lo, Wan-Yen and others},
  booktitle={Proceedings of the IEEE/CVF International Conference on Computer Vision},
  pages={4015--4026},
  year={2023}
}

@inproceedings{jang2016categorical,
  title={Categorical Reparameterization with Gumbel-Softmax},
  author={Jang, Eric and Gu, Shixiang and Poole, Ben},
  booktitle={International Conference on Learning Representations},
  year={2016}
}

@inproceedings{zhou2023zegclip,
  title={Zegclip: Towards adapting clip for zero-shot semantic segmentation},
  author={Zhou, Ziqin and Lei, Yinjie and Zhang, Bowen and Liu, Lingqiao and Liu, Yifan},
  booktitle={Proceedings of the IEEE/CVF Conference on Computer Vision and Pattern Recognition},
  pages={11175--11185},
  year={2023}
}

@inproceedings{novack2023chils,
  title={Chils: Zero-shot image classification with hierarchical label sets},
  author={Novack, Zachary and McAuley, Julian and Lipton, Zachary Chase and Garg, Saurabh},
  booktitle={International Conference on Machine Learning},
  pages={26342--26362},
  year={2023},
  organization={PMLR}
}

@inproceedings{Gu2021OpenvocabularyOD,
  title={Open-vocabulary Object Detection via Vision and Language Knowledge Distillation},
  author={Xiuye Gu and Tsung-Yi Lin and Weicheng Kuo and Yin Cui},
  booktitle={International Conference on Learning Representations},
  year={2021},
  url={https://api.semanticscholar.org/CorpusID:238744187}
}

@inproceedings{baldrati2023zero,
  title={Zero-shot composed image retrieval with textual inversion},
  author={Baldrati, Alberto and Agnolucci, Lorenzo and Bertini, Marco and Del Bimbo, Alberto},
  booktitle={Proceedings of the IEEE/CVF International Conference on Computer Vision},
  pages={15338--15347},
  year={2023}
}

@inproceedings{hendriksen2022extending,
  title={Extending CLIP for Category-to-image Retrieval in E-commerce},
  author={Hendriksen, Mariya and Bleeker, Maurits and Vakulenko, Svitlana and van Noord, Nanne and Kuiper, Ernst and de Rijke, Maarten},
  booktitle={European Conference on Information Retrieval},
  pages={289--303},
  year={2022},
  organization={Springer}
}

@inproceedings{zhang2023adding,
  title={Adding conditional control to text-to-image diffusion models},
  author={Zhang, Lvmin and Rao, Anyi and Agrawala, Maneesh},
  booktitle={Proceedings of the IEEE/CVF International Conference on Computer Vision},
  pages={3836--3847},
  year={2023}
}

@inproceedings{ruiz2023dreambooth,
  title={Dreambooth: Fine tuning text-to-image diffusion models for subject-driven generation},
  author={Ruiz, Nataniel and Li, Yuanzhen and Jampani, Varun and Pritch, Yael and Rubinstein, Michael and Aberman, Kfir},
  booktitle={Proceedings of the IEEE/CVF Conference on Computer Vision and Pattern Recognition},
  pages={22500--22510},
  year={2023}
}

@inproceedings{zhu2023tryondiffusion,
  title={Tryondiffusion: A tale of two unets},
  author={Zhu, Luyang and Yang, Dawei and Zhu, Tyler and Reda, Fitsum and Chan, William and Saharia, Chitwan and Norouzi, Mohammad and Kemelmacher-Shlizerman, Ira},
  booktitle={Proceedings of the IEEE/CVF Conference on Computer Vision and Pattern Recognition},
  pages={4606--4615},
  year={2023}
}

@inproceedings{wang2022contrastvae,
  title={Contrastvae: Contrastive variational autoencoder for sequential recommendation},
  author={Wang, Yu and Zhang, Hengrui and Liu, Zhiwei and Yang, Liangwei and Yu, Philip S},
  booktitle={Proceedings of the 31st ACM International Conference on Information \& Knowledge Management},
  pages={2056--2066},
  year={2022}
}

@article{karra2024interarec,
  title={InteraRec: Interactive Recommendations Using Multimodal Large Language Models},
  author={Karra, Saketh Reddy and Tulabandhula, Theja},
  journal={arXiv preprint arXiv:2403.00822},
  year={2024}
}

@article{liu2024rec,
  title={Rec-GPT4V: Multimodal Recommendation with Large Vision-Language Models},
  author={Liu, Yuqing and Wang, Yu and Sun, Lichao and Yu, Philip S},
  journal={arXiv preprint arXiv:2402.08670},
  year={2024}
}

@inproceedings{yan2023personalized,
  title={Personalized showcases: Generating multi-modal explanations for recommendations},
  author={Yan, An and He, Zhankui and Li, Jiacheng and Zhang, Tianyang and McAuley, Julian},
  booktitle={Proceedings of the 46th International ACM SIGIR Conference on Research and Development in Information Retrieval},
  pages={2251--2255},
  year={2023}
}

@article{team2023gemini,
  title={Gemini: a family of highly capable multimodal models},
  author={Team, Gemini and Anil, Rohan and Borgeaud, Sebastian and Wu, Yonghui and Alayrac, Jean-Baptiste and Yu, Jiahui and Soricut, Radu and Schalkwyk, Johan and Dai, Andrew M and Hauth, Anja and others},
  journal={arXiv preprint arXiv:2312.11805},
  year={2023}
}

@article{videoworldsimulators2024,
  title={Video generation models as world simulators},
  author={Tim Brooks and Bill Peebles and Connor Holmes and Will DePue and Yufei Guo and Li Jing and David Schnurr and Joe Taylor and Troy Luhman and Eric Luhman and Clarence Ng and Ricky Wang and Aditya Ramesh},
  year={2024},
  url={https://openai.com/research/video-generation-models-as-world-simulators},
}

@article{lin2023diffusion,
  title={Diffusion models for time-series applications: a survey},
  author={Lin, Lequan and Li, Zhengkun and Li, Ruikun and Li, Xuliang and Gao, Junbin},
  journal={Frontiers of Information Technology \& Electronic Engineering},
  pages={1--23},
  year={2023},
  publisher={Springer}
}

@misc{amazonARtech,
  title = {Amazon rolls out a new AR shopping feature for viewing multiple items at once},
  author = {Perez, Sarah},
  howpublished = {\url{https://techcrunch.com/2020/08/25/amazon-rolls-out-a-new-ar-shopping-feature-for-viewing-multiple-items-at-once/}},
  year = {2020},
  note = {Accessed: 2024-04-13}
}

@misc{rufus,
  title = {Amazon announces Rufus, a new generative AI-powered conversational shopping experience},
  author = {Rajiv Mehta},
  howpublished = {\url{https://www.aboutamazon.com/news/retail/amazon-rufus}},
  year= {2024},
  note = {Accessed: 2024-06-25}
}

@misc{shopwithai,
  title = {What Is Shop with Google AI and Why Should I Care?},
  author = { Selena Templeton},
  howpublished = {\url{https://www.singlegrain.com/blog/n/shop-with-google-ai/}},
  year= {2024},
  note = {Accessed: 2024-06-25}
}

@misc{amazonARcsa,
  title = {Amazon enhances online shopping experience with AI, AR},
  author = {Berthiaume, Dan},
  howpublished = {\url{https://chainstoreage.com/amazon-enhances-online-shopping-experience-ai-ar}},
  year = {2023},
  note = {Accessed: 2024-04-13}
}

@article{billinghurst2002augmented,
  title={Augmented reality in education},
  author={Billinghurst, Mark},
  journal={New horizons for learning},
  volume={12},
  number={5},
  pages={1--5},
  year={2002}
}

@article{dennler2021augmented,
  title={Augmented reality in the operating room: a clinical feasibility study},
  author={Dennler, Cyrill and Bauer, David E and Scheibler, Anne-Gita and Spirig, Jos{\'e} and G{\"o}tschi, Tobias and F{\"u}rnstahl, Philipp and Farshad, Mazda},
  journal={BMC musculoskeletal disorders},
  volume={22},
  number={1},
  pages={451},
  year={2021},
  publisher={Springer}
}

@inproceedings{reuksupasompon2018ar,
  title={Ar development for room design},
  author={Reuksupasompon, Peeranut and Aruncharathorn, Maytichai and Vittayakorn, Sirion},
  booktitle={2018 15th International Joint Conference on Computer Science and Software Engineering (JCSSE)},
  pages={1--6},
  year={2018},
  organization={IEEE}
}

@article{yuan2013mixed,
  title={A mixed reality virtual clothes try-on system},
  author={Yuan, Miaolong and Khan, Ishtiaq Rasool and Farbiz, Farzam and Yao, Susu and Niswar, Arthur and Foo, Min-Hui},
  journal={IEEE Transactions on Multimedia},
  volume={15},
  number={8},
  pages={1958--1968},
  year={2013},
  publisher={IEEE}
}

@inproceedings{han2018viton,
  title={Viton: An image-based virtual try-on network},
  author={Han, Xintong and Wu, Zuxuan and Wu, Zhe and Yu, Ruichi and Davis, Larry S},
  booktitle={Proceedings of the IEEE conference on computer vision and pattern recognition},
  pages={7543--7552},
  year={2018}
}

@misc{beauty2023compare,
  title = {L'Oréal and Amazon Just Rolled Out a Virtual Makeup Try-on Functionality},
  author = {Prinzivalli, Leah},
  howpublished = {\url{https://www.allure.com/story/loreal-amazon-modiface-virtual-try-on-lipstick}},
  year = {2019},
  note = {Accessed: 2024-04-13}
}

@inproceedings{borges2019virtual,
  title={A virtual makeup augmented reality system},
  author={Borges, Aline de F{\'a}tima Soares and Morimoto, Carlos H},
  booktitle={2019 21st Symposium on Virtual and Augmented Reality (SVR)},
  pages={34--42},
  year={2019},
  organization={IEEE}
}

@inproceedings{truong2019multimodal,
  title={Multimodal review generation for recommender systems},
  author={Truong, Quoc-Tuan and Lauw, Hady},
  booktitle={The World Wide Web Conference},
  pages={1864--1874},
  year={2019}
}

@inproceedings{shao2021controllable,
  title={Controllable and diverse text generation in e-commerce},
  author={Shao, Huajie and Wang, Jun and Lin, Haohong and Zhang, Xuezhou and Zhang, Aston and Ji, Heng and Abdelzaher, Tarek},
  booktitle={Proceedings of the Web Conference 2021},
  pages={2392--2401},
  year={2021}
}

@misc{amazon2023reviews,
  title = {How Amazon continues to improve the customer reviews experience with generative AI},
  author = {Schermerhorn, Vaughn},
  howpublished = {\url{https://www.aboutamazon.com/news/amazon-ai/amazon-improves-customer-reviews-with-generative-ai}},
  year = {2023},
  note = {Accessed: 2024-04-13}
}

@inproceedings{corneanu2024latentpaint,
  title={LatentPaint: Image Inpainting in Latent Space With Diffusion Models},
  author={Corneanu, Ciprian and Gadde, Raghudeep and Martinez, Aleix M},
  booktitle={Proceedings of the IEEE/CVF Winter Conference on Applications of Computer Vision},
  pages={4334--4343},
  year={2024}
}

@article{deng2022toward,
  title={Toward personalized answer generation in e-commerce via multi-perspective preference modeling},
  author={Deng, Yang and Li, Yaliang and Zhang, Wenxuan and Ding, Bolin and Lam, Wai},
  journal={ACM Transactions on Information Systems (TOIS)},
  volume={40},
  number={4},
  pages={1--28},
  year={2022},
  publisher={ACM New York, NY}
}

@inproceedings{xiao2022abstract,
  title={From abstract to details: A generative multimodal fusion framework for recommendation},
  author={Xiao, Fangxiong and Deng, Lixi and Chen, Jingjing and Ji, Houye and Yang, Xiaorui and Ding, Zhuoye and Long, Bo},
  booktitle={Proceedings of the 30th ACM International Conference on Multimedia},
  pages={258--267},
  year={2022}
}

@inproceedings{novgorodov2019generating,
  title={Generating product descriptions from user reviews},
  author={Novgorodov, Slava and Guy, Ido and Elad, Guy and Radinsky, Kira},
  booktitle={The world wide web conference},
  pages={1354--1364},
  year={2019}
}

@article{wang2024stablegarment,
  title={StableGarment: Garment-Centric Generation via Stable Diffusion},
  author={Wang, Rui and Guo, Hailong and Liu, Jiaming and Li, Huaxia and Zhao, Haibo and Tang, Xu and Hu, Yao and Tang, Hao and Li, Peipei},
  journal={arXiv preprint arXiv:2403.10783},
  year={2024}
}

@article{xu2024ootdiffusion,
  title={OOTDiffusion: Outfitting Fusion based Latent Diffusion for Controllable Virtual Try-on},
  author={Xu, Yuhao and Gu, Tao and Chen, Weifeng and Chen, Chengcai},
  journal={arXiv preprint arXiv:2403.01779},
  year={2024}
}

@article{xu2024difashion,
  title={DiFashion: Towards Personalized Outfit Generation},
  author={Xu, Yiyan and Wang, Wenjie and Feng, Fuli and Ma, Yunshan and Zhang, Jizhi and He, Xiangnan},
  journal={arXiv preprint arXiv:2402.17279},
  year={2024}
}

@inproceedings{wang2024texfit,
  title={TexFit: Text-Driven Fashion Image Editing with Diffusion Models},
  author={Wang, Tongxin and Ye, Mang},
  booktitle={Proceedings of the AAAI Conference on Artificial Intelligence},
  volume={38},
  number={9},
  pages={10198--10206},
  year={2024}
}

@inproceedings{wei2022towards,
  title={Towards personalized bundle creative generation with contrastive non-autoregressive decoding},
  author={Wei, Penghui and Liu, Shaoguo and Yang, Xuanhua and Wang, Liang and Zheng, Bo},
  booktitle={Proceedings of the 45th International ACM SIGIR Conference on Research and Development in Information Retrieval},
  pages={2634--2638},
  year={2022}
}

@article{wang2023generate,
  title={Generate E-commerce Product Background by Integrating Category Commonality and Personalized Style},
  author={Wang, Haohan and Feng, Wei and Lu, Yang and Li, Yaoyu and Zhang, Zheng and Lv, Jingjing and Zhu, Xin and Shen, Junjie and Lin, Zhangang and Bo, Lixing and others},
  journal={arXiv preprint arXiv:2312.13309},
  year={2023}
}

@inproceedings{chen2021automated,
  title={Automated creative optimization for e-commerce advertising},
  author={Chen, Jin and Xu, Ju and Jiang, Gangwei and Ge, Tiezheng and Zhang, Zhiqiang and Lian, Defu and Zheng, Kai},
  booktitle={Proceedings of the Web Conference 2021},
  pages={2304--2313},
  year={2021}
}

@article{shilova2023adbooster,
  title={AdBooster: Personalized Ad Creative Generation using Stable Diffusion Outpainting},
  author={Shilova, Veronika and Santos, Ludovic Dos and Vasile, Flavian and Racic, Ga{\"e}tan and Tanielian, Ugo},
  journal={arXiv preprint arXiv:2309.11507},
  year={2023}
}

@misc{seyfioglu2024diffuse,
      title={Diffuse to Choose: Enriching Image Conditioned Inpainting in Latent Diffusion Models for Virtual Try-All}, 
      author={Mehmet Saygin Seyfioglu and Karim Bouyarmane and Suren Kumar and Amir Tavanaei and Ismail B. Tutar},
      year={2024},
      eprint={2401.13795},
      archivePrefix={arXiv},
      primaryClass={cs.CV}
}

@inproceedings{jones2023learning,
author = {Jones, Rosie},
title = {Learning to Understand Audio and Multimodal Content},
year = {2023},
isbn = {9781450394079},
publisher = {Association for Computing Machinery},
address = {New York, NY, USA},
url = {https://doi.org/10.1145/3539597.3572333},
doi = {10.1145/3539597.3572333},
booktitle = {Proceedings of the Sixteenth ACM International Conference on Web Search and Data Mining},
pages = {4–5},
numpages = {2},
location = {, Singapore, Singapore, },
series = {WSDM '23}
}

@inproceedings{lei20mmmicrov,
author = {Lei, Chenyi and Liu, Yong and Zhang, Lingzi and Wang, Guoxin and Tang, Haihong and Li, Houqiang and Miao, Chunyan},
title = {SEMI: A Sequential Multi-Modal Information Transfer Network for E-Commerce Micro-Video Recommendations},
year = {2021},
isbn = {9781450383325},
publisher = {Association for Computing Machinery},
address = {New York, NY, USA},
url = {https://doi.org/10.1145/3447548.3467189},
doi = {10.1145/3447548.3467189},
booktitle = {Proceedings of the 27th ACM SIGKDD Conference on Knowledge Discovery \& Data Mining},
pages = {3161–3171},
numpages = {11},
location = {Virtual Event, Singapore},
series = {KDD '21}
}

@inproceedings{wei2019mmgcn,
  title={MMGCN: Multi-modal graph convolution network for personalized recommendation of micro-video},
  author={Wei, Yinwei and Wang, Xiang and Nie, Liqiang and He, Xiangnan and Hong, Richang and Chua, Tat-Seng},
  booktitle={Proceedings of the 27th ACM international conference on multimedia},
  pages={1437--1445},
  year={2019}
}

@inproceedings{yi2022multi,
  title={Multi-modal graph contrastive learning for micro-video recommendation},
  author={Yi, Zixuan and Wang, Xi and Ounis, Iadh and Macdonald, Craig},
  booktitle={Proceedings of the 45th International ACM SIGIR Conference on Research and Development in Information Retrieval},
  pages={1807--1811},
  year={2022}
}

@inproceedings{chen2021learning,
  title={Learning audio embeddings with user listening data for content-based music recommendation},
  author={Chen, Ke and Liang, Beici and Ma, Xiaoshuan and Gu, Minwei},
  booktitle={ICASSP 2021-2021 IEEE International Conference on Acoustics, Speech and Signal Processing (ICASSP)},
  pages={3015--3019},
  year={2021},
  organization={IEEE}
}

@inproceedings{huang2020large,
  title={Large-scale weakly-supervised content embeddings for music recommendation and tagging},
  author={Huang, Qingqing and Jansen, Aren and Zhang, Li and Ellis, Daniel PW and Saurous, Rif A and Anderson, John},
  booktitle={ICASSP 2020-2020 IEEE International Conference on Acoustics, Speech and Signal Processing (ICASSP)},
  pages={8364--8368},
  year={2020},
  organization={IEEE}
}

@article{deldjoo2024content,
  title={Content-driven music recommendation: Evolution, state of the art, and challenges},
  author={Deldjoo, Yashar and Schedl, Markus and Knees, Peter},
  journal={Computer Science Review},
  volume={51},
  pages={100618},
  year={2024},
  publisher={Elsevier}
}

@article{sun2022long,
  title={Long-form video-language pre-training with multimodal temporal contrastive learning},
  author={Sun, Yuchong and Xue, Hongwei and Song, Ruihua and Liu, Bei and Yang, Huan and Fu, Jianlong},
  journal={Advances in neural information processing systems},
  volume={35},
  pages={38032--38045},
  year={2022}
}

@article{vyas2023audiobox,
  title={Audiobox: Unified audio generation with natural language prompts},
  author={Vyas, Apoorv and Shi, Bowen and Le, Matthew and Tjandra, Andros and Wu, Yi-Chiao and Guo, Baishan and Zhang, Jiemin and Zhang, Xinyue and Adkins, Robert and Ngan, William and others},
  journal={arXiv preprint arXiv:2312.15821},
  year={2023}
}

@article{briot2017deep,
  title={Deep learning techniques for music generation--a survey},
  author={Briot, Jean-Pierre and Hadjeres, Ga{\"e}tan and Pachet, Fran{\c{c}}ois-David},
  journal={arXiv preprint arXiv:1709.01620},
  year={2017}
}

@article{lam2024efficient,
  title={Efficient neural music generation},
  author={Lam, Max WY and Tian, Qiao and Li, Tang and Yin, Zongyu and Feng, Siyuan and Tu, Ming and Ji, Yuliang and Xia, Rui and Ma, Mingbo and Song, Xuchen and others},
  journal={Advances in Neural Information Processing Systems},
  volume={36},
  year={2024}
}

@article{liu2023ai,
  title={Ai-empowered persuasive video generation: A survey},
  author={Liu, Chang and Yu, Han},
  journal={ACM Computing Surveys},
  volume={55},
  number={13s},
  pages={1--31},
  year={2023},
  publisher={ACM New York, NY}
}

@inproceedings{loukili23marketing,
author = {Loukili, Soumaya and Fennan, Abdelhadi and Elaachak, Lotfi},
title = {Applications of Text Generation in Digital Marketing: a review},
year = {2023},
isbn = {9798400700194},
publisher = {Association for Computing Machinery},
address = {New York, NY, USA},
url = {https://doi.org/10.1145/3607720.3608451},
doi = {10.1145/3607720.3608451},
booktitle = {Proceedings of the 6th International Conference on Networking, Intelligent Systems \& Security},
articleno = {69},
numpages = {8},
location = {, Larache, Morocco, },
series = {NISS '23}
}

@article{mayahi2022impact,
  title={The impact of generative ai on the future of visual content marketing},
  author={Mayahi, Shiva and Vidrih, Marko},
  journal={arXiv preprint arXiv:2211.12660},
  year={2022}
}

@article{dhariwal2020jukebox,
  title={Jukebox: A generative model for music},
  author={Dhariwal, Prafulla and Jun, Heewoo and Payne, Christine and Kim, Jong Wook and Radford, Alec and Sutskever, Ilya},
  journal={arXiv preprint arXiv:2005.00341},
  year={2020}
}

@inproceedings{chen16infogan,
author = {Chen, Xi and Duan, Yan and Houthooft, Rein and Schulman, John and Sutskever, Ilya and Abbeel, Pieter},
title = {InfoGAN: interpretable representation learning by information maximizing generative adversarial nets},
year = {2016},
isbn = {9781510838819},
publisher = {Curran Associates Inc.},
address = {Red Hook, NY, USA},
booktitle = {Proceedings of the 30th International Conference on Neural Information Processing Systems},
pages = {2180–2188},
numpages = {9},
location = {Barcelona, Spain},
series = {NIPS'16}
}

@INPROCEEDINGS{isola17pix2pix,
  author={Isola, Phillip and Zhu, Jun-Yan and Zhou, Tinghui and Efros, Alexei A.},
  booktitle={2017 IEEE Conference on Computer Vision and Pattern Recognition (CVPR)}, 
  title={Image-to-Image Translation with Conditional Adversarial Networks}, 
  year={2017},
  volume={},
  number={},
  pages={5967-5976},
  doi={10.1109/CVPR.2017.632}
}

@article{MirzaO14conditional,
  author       = {Mehdi Mirza and
                  Simon Osindero},
  title        = {Conditional Generative Adversarial Nets},
  journal      = {CoRR},
  volume       = {abs/1411.1784},
  year         = {2014},
  url          = {http://arxiv.org/abs/1411.1784},
  eprinttype    = {arXiv},
  eprint       = {1411.1784},
  bibsource    = {dblp computer science bibliography, https://dblp.org}
}

@article{Su2018GANQPAN,
  title={GAN-QP: A Novel GAN Framework without Gradient Vanishing and Lipschitz Constraint},
  author={Jianlin Su},
  journal={ArXiv},
  year={2018},
  volume={abs/1811.07296},
  url={https://api.semanticscholar.org/CorpusID:53713813}
}

@article{Chakraborty_2024,
doi = {10.1088/2632-2153/ad1f77},
url = {https://dx.doi.org/10.1088/2632-2153/ad1f77},
year = {2024},
month = {jan},
publisher = {IOP Publishing},
volume = {5},
number = {1},
pages = {011001},
author = {Tanujit Chakraborty and Ujjwal Reddy K S and Shraddha M Naik and Madhurima Panja and Bayapureddy Manvitha},
title = {Ten years of generative adversarial nets (GANs): a survey of the state-of-the-art},
journal = {Machine Learning: Science and Technology}
}

@inproceedings{arjovsky17waserstein,
    author = {Arjovsky, Martin and Chintala, Soumith and Bottou, L\'{e}on},
    title = {Wasserstein generative adversarial networks},
    year = {2017},
    publisher = {JMLR.org},
    booktitle = {Proceedings of the 34th International Conference on Machine Learning - Volume 70},
    pages = {214–223},
    numpages = {10},
    location = {Sydney, NSW, Australia},
    series = {ICML'17}
}

@inproceedings{zhang19vanishing,
author = {Zhang, Zhaoyu and Luo, Changwei and Yu, Jun},
title = {Towards the Gradient Vanishing, Divergence Mismatching and Mode Collapse of Generative Adversarial Nets},
year = {2019},
isbn = {9781450369763},
publisher = {Association for Computing Machinery},
address = {New York, NY, USA},
url = {https://doi.org/10.1145/3357384.3358081},
doi = {10.1145/3357384.3358081},
booktitle = {Proceedings of the 28th ACM International Conference on Information and Knowledge Management},
pages = {2377–2380},
numpages = {4},
location = {Beijing, China},
series = {CIKM '19}
}

@inproceedings{huang2022multimodal,
  title={Multimodal conditional image synthesis with product-of-experts gans},
  author={Huang, Xun and Mallya, Arun and Wang, Ting-Chun and Liu, Ming-Yu},
  booktitle={European Conference on Computer Vision},
  pages={91--109},
  year={2022},
  organization={Springer}
}

@inproceedings{ziegler2022multi,
  title={Multi-modal conditional GAN: Data synthesis in the medical domain},
  author={Ziegler, Jonathan David and Subramaniam, Sajanth and Azzarito, Michela and Doyle, Orla and Krusche, Peter and Coroller, Thibaud},
  booktitle={NeurIPS 2022 Workshop on Synthetic Data for Empowering ML Research},
  year={2022}
}

@article{gao2021recommender,
  title={Recommender systems based on generative adversarial networks: A problem-driven perspective},
  author={Gao, Min and Zhang, Junwei and Yu, Junliang and Li, Jundong and Wen, Junhao and Xiong, Qingyu},
  journal={Information Sciences},
  volume={546},
  pages={1166--1185},
  year={2021},
  publisher={Elsevier}
}

@article{liu2021clothing,
  title={Clothing generation by multi-modal embedding: A compatibility matrix-regularized GAN model},
  author={Liu, Linlin and Zhang, Haijun and Zhou, Dongliang},
  journal={Image and Vision Computing},
  volume={107},
  pages={104097},
  year={2021},
  publisher={Elsevier}
}

@article{pandey2020poly,
  title={Poly-GAN: Multi-conditioned GAN for fashion synthesis},
  author={Pandey, Nilesh and Savakis, Andreas},
  journal={Neurocomputing},
  volume={414},
  pages={356--364},
  year={2020},
  publisher={Elsevier}
}

@article{zhou2023bc,
  title={BC-GAN: A Generative Adversarial Network for Synthesizing a Batch of Collocated Clothing},
  author={Zhou, Dongliang and Zhang, Haijun and Ma, Jianghong and Shi, Jianyang},
  journal={IEEE Transactions on Circuits and Systems for Video Technology},
  year={2023},
  publisher={IEEE}
}

@article{liu2019swapgan,
  title={SwapGAN: A multistage generative approach for person-to-person fashion style transfer},
  author={Liu, Yu and Chen, Wei and Liu, Li and Lew, Michael S},
  journal={IEEE Transactions on Multimedia},
  volume={21},
  number={9},
  pages={2209--2222},
  year={2019},
  publisher={IEEE}
}

@article{tautkute2021want,
  title={I want this product but different: Multimodal retrieval with synthetic query expansion},
  author={Tautkute, Ivona and Trzcinski, Tomasz},
  journal={arXiv preprint arXiv:2102.08871},
  year={2021}
}

@inproceedings{lin2023contrastive,
  title={Contrastive Intra-and Inter-Modality Generation for Enhancing Incomplete Multimedia Recommendation},
  author={Lin, Zhenghong and Tan, Yanchao and Zhan, Yunfei and Liu, Weiming and Wang, Fan and Chen, Chaochao and Wang, Shiping and Yang, Carl},
  booktitle={Proceedings of the 31st ACM International Conference on Multimedia},
  pages={6234--6242},
  year={2023}
}

@article{brown2020language,
  title={Language models are few-shot learners},
  author={Brown, Tom and Mann, Benjamin and Ryder, Nick and Subbiah, Melanie and Kaplan, Jared D and Dhariwal, Prafulla and Neelakantan, Arvind and Shyam, Pranav and Sastry, Girish and Askell, Amanda and others},
  journal={Advances in neural information processing systems},
  volume={33},
  pages={1877--1901},
  year={2020}
}

@article{wu2023survey,
  title={A survey on large language models for recommendation},
  author={Wu, Likang and Zheng, Zhi and Qiu, Zhaopeng and Wang, Hao and Gu, Hongchao and Shen, Tingjia and Qin, Chuan and Zhu, Chen and Zhu, Hengshu and Liu, Qi and others},
  journal={arXiv preprint arXiv:2305.19860},
  year={2023}
}

@article{hu2021lora,
  title={Lora: Low-rank adaptation of large language models},
  author={Hu, Edward J and Shen, Yelong and Wallis, Phillip and Allen-Zhu, Zeyuan and Li, Yuanzhi and Wang, Shean and Wang, Lu and Chen, Weizhu},
  journal={arXiv preprint arXiv:2106.09685},
  year={2021}
}

@article{chen2015microsoft,
  title={Microsoft coco captions: Data collection and evaluation server},
  author={Chen, Xinlei and Fang, Hao and Lin, Tsung-Yi and Vedantam, Ramakrishna and Gupta, Saurabh and Doll{\'a}r, Piotr and Zitnick, C Lawrence},
  journal={arXiv preprint arXiv:1504.00325},
  year={2015}
}

@inproceedings{li2023blip,
  title={Blip-2: Bootstrapping language-image pre-training with frozen image encoders and large language models},
  author={Li, Junnan and Li, Dongxu and Savarese, Silvio and Hoi, Steven},
  booktitle={International conference on machine learning},
  pages={19730--19742},
  year={2023},
  organization={PMLR}
}

@article{alayrac2022flamingo,
  title={Flamingo: a visual language model for few-shot learning},
  author={Alayrac, Jean-Baptiste and Donahue, Jeff and Luc, Pauline and Miech, Antoine and Barr, Iain and Hasson, Yana and Lenc, Karel and Mensch, Arthur and Millican, Katherine and Reynolds, Malcolm and others},
  journal={Advances in neural information processing systems},
  volume={35},
  pages={23716--23736},
  year={2022}
}

@article{kong2024audio,
  title={Audio Flamingo: A Novel Audio Language Model with Few-Shot Learning and Dialogue Abilities},
  author={Kong, Zhifeng and Goel, Arushi and Badlani, Rohan and Ping, Wei and Valle, Rafael and Catanzaro, Bryan},
  journal={arXiv preprint arXiv:2402.01831},
  year={2024}
}

@article{deshmukh2023pengi,
  title={Pengi: An audio language model for audio tasks},
  author={Deshmukh, Soham and Elizalde, Benjamin and Singh, Rita and Wang, Huaming},
  journal={Advances in Neural Information Processing Systems},
  volume={36},
  pages={18090--18108},
  year={2023}
}

@article{han2023imagebind,
  title={Imagebind-llm: Multi-modality instruction tuning},
  author={Han, Jiaming and Zhang, Renrui and Shao, Wenqi and Gao, Peng and Xu, Peng and Xiao, Han and Zhang, Kaipeng and Liu, Chris and Wen, Song and Guo, Ziyu and others},
  journal={arXiv preprint arXiv:2309.03905},
  year={2023}
}

@article{moon2023anymal,
  title={Anymal: An efficient and scalable any-modality augmented language model},
  author={Moon, Seungwhan and Madotto, Andrea and Lin, Zhaojiang and Nagarajan, Tushar and Smith, Matt and Jain, Shashank and Yeh, Chun-Fu and Murugesan, Prakash and Heidari, Peyman and Liu, Yue and others},
  journal={arXiv preprint arXiv:2309.16058},
  year={2023}
}

@article{xie2024travelplanner,
  title={Travelplanner: A benchmark for real-world planning with language agents},
  author={Xie, Jian and Zhang, Kai and Chen, Jiangjie and Zhu, Tinghui and Lou, Renze and Tian, Yuandong and Xiao, Yanghua and Su, Yu},
  journal={arXiv preprint arXiv:2402.01622},
  year={2024}
}

@inproceedings{zhu2024consistent,
  title={Consistent Multimodal Generation via A Unified GAN Framework},
  author={Zhu, Zhen and Li, Yijun and Lyu, Weijie and Singh, Krishna Kumar and Shu, Zhixin and Pirk, S{\"o}ren and Hoiem, Derek},
  booktitle={Proceedings of the IEEE/CVF Winter Conference on Applications of Computer Vision},
  pages={5048--5057},
  year={2024}
}

@inproceedings{ronneberger2015u,
  title={U-net: Convolutional networks for biomedical image segmentation},
  author={Ronneberger, Olaf and Fischer, Philipp and Brox, Thomas},
  booktitle={Medical image computing and computer-assisted intervention--MICCAI 2015: 18th international conference, Munich, Germany, October 5-9, 2015, proceedings, part III 18},
  pages={234--241},
  year={2015},
  organization={Springer}
}

@article{cao19openpose,
  author = {Z. {Cao} and G. {Hidalgo Martinez} and T. {Simon} and S. {Wei} and Y. A. {Sheikh}},
  journal = {IEEE Transactions on Pattern Analysis and Machine Intelligence},
  title = {OpenPose: Realtime Multi-Person 2D Pose Estimation using Part Affinity Fields},
  year = {2019}
}

@article{Chen2023BeyondSL,
  title={Beyond Semantics: Learning a Behavior Augmented Relevance Model with Self-supervised Learning},
  author={Ze-jie Chen and Wei Chen and Jia Xu and Zhongyi Liu and Wei Zhang},
  journal={Proceedings of the 32nd ACM International Conference on Information and Knowledge Management},
  year={2023},
  url={https://api.semanticscholar.org/CorpusID:260775754}
}

@article{huang2021cross,
  title={Cross-modal contrastive learning of representations for navigation using lightweight, low-cost millimeter wave radar for adverse environmental conditions},
  author={Huang, Jui-Te and Lu, Chen-Lung and Chang, Po-Kai and Huang, Ching-I and Hsu, Chao-Chun and Huang, Po-Jui and Wang, Hsueh-Cheng and others},
  journal={IEEE Robotics and Automation Letters},
  volume={6},
  number={2},
  pages={3333--3340},
  year={2021},
  publisher={IEEE}
}

@inproceedings{zhang2021contrastive,
  title={Contrastive self-supervised learning for text-independent speaker verification},
  author={Zhang, Haoran and Zou, Yuexian and Wang, Helin},
  booktitle={ICASSP 2021-2021 IEEE International Conference on Acoustics, Speech and Signal Processing (ICASSP)},
  pages={6713--6717},
  year={2021},
  organization={IEEE}
}

@inproceedings{bansal2023cleanclip,
  title={Cleanclip: Mitigating data poisoning attacks in multimodal contrastive learning},
  author={Bansal, Hritik and Singhi, Nishad and Yang, Yu and Yin, Fan and Grover, Aditya and Chang, Kai-Wei},
  booktitle={Proceedings of the IEEE/CVF International Conference on Computer Vision},
  pages={112--123},
  year={2023}
}

@article{yu2023devil,
  title={The devil is in the details: A deep dive into the rabbit hole of data filtering},
  author={Yu, Haichao and Tian, Yu and Kumar, Sateesh and Yang, Linjie and Wang, Heng},
  journal={arXiv preprint arXiv:2309.15954},
  year={2023}
}

@article{schuhmann2022laion,
  title={Laion-5b: An open large-scale dataset for training next generation image-text models},
  author={Schuhmann, Christoph and Beaumont, Romain and Vencu, Richard and Gordon, Cade and Wightman, Ross and Cherti, Mehdi and Coombes, Theo and Katta, Aarush and Mullis, Clayton and Wortsman, Mitchell and others},
  journal={Advances in Neural Information Processing Systems},
  volume={35},
  pages={25278--25294},
  year={2022}
}

@inproceedings{ruan2023mm,
  title={Mm-diffusion: Learning multi-modal diffusion models for joint audio and video generation},
  author={Ruan, Ludan and Ma, Yiyang and Yang, Huan and He, Huiguo and Liu, Bei and Fu, Jianlong and Yuan, Nicholas Jing and Jin, Qin and Guo, Baining},
  booktitle={Proceedings of the IEEE/CVF Conference on Computer Vision and Pattern Recognition},
  pages={10219--10228},
  year={2023}
}

@article{croitoru2023diffusion,
  title={Diffusion models in vision: A survey},
  author={Croitoru, Florinel-Alin and Hondru, Vlad and Ionescu, Radu Tudor and Shah, Mubarak},
  journal={IEEE Transactions on Pattern Analysis and Machine Intelligence},
  year={2023},
  publisher={IEEE}
}

@inproceedings{haokai24mmcdiff,
author = {Ma, Haokai and Yang, Yimeng and Meng, Lei and Xie, Ruobing and Meng, Xiangxu},
title = {Multimodal Conditioned Diffusion Model for Recommendation},
year = {2024},
isbn = {9798400701726},
publisher = {Association for Computing Machinery},
address = {New York, NY, USA},
url = {https://doi.org/10.1145/3589335.3651956},
doi = {10.1145/3589335.3651956},
booktitle = {Companion Proceedings of the ACM on Web Conference 2024},
pages = {1733–1740},
numpages = {8},
location = {, Singapore, Singapore, },
series = {WWW '24}
}

@InProceedings{balanced_vqa_v2,
author = {Yash Goyal and Tejas Khot and Douglas Summers{-}Stay and Dhruv Batra and Devi Parikh},
title = {Making the {V} in {VQA} Matter: Elevating the Role of Image Understanding in {V}isual {Q}uestion {A}nswering},
booktitle = {Conference on Computer Vision and Pattern Recognition (CVPR)},
year = {2017},
}

@misc{anthropic2024claude,
  title = {Introducing the next generation of Claude},
  author = {Anthropic Team},
  howpublished = {\url{https://www.anthropic.com/news/claude-3-family}},
  year = {2024},
  note = {Accessed: 2024-05-15}
}

@inproceedings{liang2018variational,
  title={Variational autoencoders for collaborative filtering},
  author={Liang, Dawen and Krishnan, Rahul G and Hoffman, Matthew D and Jebara, Tony},
  booktitle={Proceedings of the 2018 world wide web conference},
  pages={689--698},
  year={2018}
}

@Inproceedings{Chen2024mmdiff,
 author = {Changyou Chen and Han Ding and Bunyamin Sisman and Yi Xu and Ouye Xie and Benjamin Yao and Son Tran and Belinda Zeng},
 title = {Diffusion models for multi-modal generative modeling},
 year = {2024},
 url = {https://www.amazon.science/publications/diffusion-models-for-multi-modal-generative-modeling},
 booktitle = {ICLR 2024},
}

@article{zihao23diffurec,
author = {Li, Zihao and Sun, Aixin and Li, Chenliang},
year = {2023},
month = {12},
pages = {1-28},
title = {DiffuRec: A Diffusion Model for Sequential Recommendation},
volume = {42},
journal = {ACM Transactions on Information Systems},
doi = {10.1145/3631116}
}

@article{sevegnani2022contrastive,
  title={Contrastive learning for interactive recommendation in fashion},
  author={Sevegnani, Karin and Seshadri, Arjun and Wang, Tian and Beniwal, Anurag and McAuley, Julian and Lu, Alan and Medioni, Gerard},
  journal={arXiv preprint arXiv:2207.12033},
  year={2022}
}

@article{alpay2023multimodal,
  title={Multimodal video retrieval with CLIP: a user study},
  author={Alpay, Tayfun and Magg, Sven and Broze, Philipp and Speck, Daniel},
  journal={Information Retrieval Journal},
  volume={26},
  number={1-2},
  year={2023},
  publisher={Springer Netherlands Dordrecht}
}

@article{jiang2024diffmm,
  title={DiffMM: Multi-Modal Diffusion Model for Recommendation},
  author={Jiang, Yangqin and Xia, Lianghao and Wei, Wei and Luo, Da and Lin, Kangyi and Huang, Chao},
  journal={arXiv preprint arXiv:2406.11781},
  year={2024}
}

@inproceedings{chattopadhyay2023learning,
  title={Learning graph variational autoencoders with constraints and structured priors for conditional indoor 3d scene generation},
  author={Chattopadhyay, Aditya and Zhang, Xi and Wipf, David Paul and Arora, Himanshu and Vidal, Ren{\'e}},
  booktitle={Proceedings of the IEEE/CVF winter conference on applications of computer vision},
  pages={785--794},
  year={2023}
}

@article{zhou2024disentangled,
  title={Disentangled Graph Variational Auto-Encoder for Multimodal Recommendation With Interpretability},
  author={Zhou, Xin and Miao, Chunyan},
  journal={IEEE Transactions on Multimedia},
  year={2024},
  publisher={IEEE}
}

@inproceedings{yang2022vision,
  title={Vision-language pre-training with triple contrastive learning},
  author={Yang, Jinyu and Duan, Jiali and Tran, Son and Xu, Yi and Chanda, Sampath and Chen, Liqun and Zeng, Belinda and Chilimbi, Trishul and Huang, Junzhou},
  booktitle={Proceedings of the IEEE/CVF Conference on Computer Vision and Pattern Recognition},
  pages={15671--15680},
  year={2022}
}

@inproceedings{du2022amazon,
  title={Amazon shop the look: A visual search system for fashion and home},
  author={Du, Ming and Ramisa, Arnau and KC, Amit Kumar and Chanda, Sampath and Wang, Mengjiao and Rajesh, Neelakandan and Li, Shasha and Hu, Yingchuan and Zhou, Tao and Lakshminarayana, Nagashri and others},
  booktitle={Proceedings of the 28th ACM SIGKDD Conference on Knowledge Discovery and Data Mining},
  pages={2822--2830},
  year={2022}
}

@article{kipf2016variational,
  title={Variational graph auto-encoders},
  author={Kipf, Thomas N and Welling, Max},
  journal={arXiv preprint arXiv:1611.07308},
  year={2016}
}

\end{document}